\documentclass[prd,nofootinbib,onecolumn]{revtex4}

\pdfoutput=1

\usepackage{amssymb}
\usepackage{amsmath}
\usepackage{graphicx,subfigure}
\usepackage{longtable}
\usepackage{verbatim}
\usepackage{amsfonts}

\usepackage{color}

\usepackage{graphicx,amsmath,amssymb,epsfig,verbatim,mathrsfs,array,layout,textcomp,latexsym,xspace}
\usepackage{multirow}

\allowdisplaybreaks[1]

\renewcommand{\{}{\left\lbrace}
\renewcommand{\}}{\right\rbrace}
\renewcommand{\[}{\left\lbrack}
\renewcommand{\]}{\right\rbrack}

\newcommand{\dd}[1][{}]{\mathrm{d}^{#1}\!\!\;}

\newcommand{\nn}{\nonumber}

\def \re{\textrm{Re}}
\def \im{\textrm{Im}}

\def \azeL{{H_0^L}}
\def \azeR{{H_0^R}}
\def \apaL{{H_\parallel^L}}

\def \apeL{{H_\perp^L}}

\newcommand{\gev}{\ensuremath{\mathrm{\,Ge\kern -0.1em V}}\xspace}
\newcommand{\gevcc}{\ensuremath{{\mathrm{\,Ge\kern -0.1em V\!/}c^2}}\xspace}

\newcommand{\BKpill}{$\bar B\to \bar K\pi\ell\ell$}
\newcommand{\BsKKll}{$\bar B_s\to \bar KK\ell\ell$}
\begin{document}

\begin{flushright}
DO-TH 14/10 \\
QFET-2014-09
\end{flushright}

\title{The $\boldsymbol{\bar B \to \bar K \pi \ell \ell}$ and $\boldsymbol{\bar B_s  \to \bar K K  \ell \ell}$ distributions at low hadronic recoil}

\author{Diganta Das}
\author{Gudrun Hiller}
\author{Martin Jung}
\author{Alex Shires}
\affiliation{Institut f\"ur Physik, Technische Universit\"at Dortmund, D-44221 Dortmund, Germany}

\begin{abstract}
The rare multi-body decays $\bar B \to \bar K \pi \ell \ell$ 
and $\bar B_s  \to \bar K K  \ell \ell$ are both important as backgrounds to precision  
analyses in the benchmark modes
$\bar B \to \bar K^*\ell \ell$ and $\bar B_s  \to \phi \ell \ell$
as well as sensitive probes of flavor physics in and beyond the
standard model.
We work out non-resonant  contributions to $\bar B \to \bar K \pi \ell  
\ell$ and $\bar B_s  \to \bar K K  \ell \ell$ amplitudes, 
where $\ell=e, \mu$, at low  
hadronic recoil in a model-independent way.
Using the operator product expansion in $1/m_b$, we present  
expressions for the full angular distribution. The latter allows to  
probe new combinations of
$|\Delta B|=|\Delta S|=1$  couplings and gives access to strong phases  
between non-resonant and resonant contributions.
Exact endpoint relations between transversity amplitudes based on  
Lorentz invariance are obtained.
Several phenomenological distributions including those from the angular projections to the S-, P-, D-waves are given.
Standard model branching ratios for non-resonant $\bar B \to \bar K  
\pi \ell \ell$ and $\bar B_s  \to \bar K K  \ell \ell$ decays are found to be  
in the few $10^{-8}$ region, but drop significantly if cuts around the  
$K^*$ or $\phi$ mass are employed. 
Nevertheless, the non-resonant contributions to $\bar B \to \bar K  
\pi \ell \ell$ provide the dominant background in the $\bar B \to \bar K^* \ell \ell$ 
signal region with respect to the low mass scalars.
In $\bar B_s  \to \bar K K  \ell \ell$, the narrowness of the $\phi$ allows  
for more efficient background control.
We briefly discuss lepton-flavor non-universal effects, also in view of the recent data on $R_K$.
\end{abstract}

\maketitle

\section{Introduction}

The semi-leptonic decays $\bar B \to \bar K \pi \ell \ell$ and $\bar B_s  \to \bar K K  \ell \ell$  are $|\Delta B|=|\Delta S|=1$  flavor-changing neutral current processes (FCNCs) which are sensitive to flavor physics in and beyond the standard model. 
While being phase space suppressed with respect to the corresponding resonant channels $\bar B \to \bar K^{(*)}(\phi) \ell \ell$, the non-resonant decays become important with high statistics available for future experimental analyses~\cite{Abe:2010gxa,Bediaga:2012py}. 
In particular non-resonant decays constitute important backgrounds to the forthcoming precision studies of $\bar B \to \bar K^*( \to \bar K \pi) \ell \ell$ and  $\bar B_s \to \phi( \to \bar K K) \ell \ell$ decays. 

In this work we calculate the non-resonant $\bar B \to \bar K \pi \ell \ell$ and related $\bar B_s  \to \bar K K  \ell \ell$ contributions.
The non-resonant heavy to light  decays factorize at low recoil by means of the hard scale of order of the $b-$quark mass, $m_b$, for the dilepton invariant mass squared, $q^2={\cal{O}}(m_b^2)$ \cite{Buchalla:1998mt}. 
Specifically we employ the operator product expansion (OPE) in $1/m_b$ as  put forward in Ref.~\cite{Grinstein:2004vb} (for later work, see \cite{Beylich:2011aq}) with subsequent detailed analyses for  resonant decays \cite{Bobeth:2010wg,Bobeth:2012vn}.
Requisite hadronic $\bar B \to \bar K \pi$ form factors are available from  heavy hadron chiral perturbation theory (HH$\chi$PT), {\it e.g.,}  Refs.~\cite{BD92,Lee:1992ih}, valid in the region where both 3-momenta of the final-state pseudoscalars are soft in the $B$-meson rest frame.
We perform phenomenological studies  in this region of low recoil.
In the large recoil region, a recent study is Ref.~\cite{Doring:2013wka}. Recent activities
covering semi-leptonic $b \to u \ell \nu$ tranisitions include
\cite{Faller:2013dwa,Kang:2013jaa}.

Other backgrounds  to $\bar B \to \bar K^{*} (\to \bar K \pi)  \ell \ell$ previously considered are resonant S-wave contributions from the scalar mesons $K_0^*(1430)$ or $\kappa(800)$ \cite{Becirevic:2012dp,Matias:2012qz,Blake:2012mb}.
Here we discuss additionally features of the $D$-wave background. 
While heavier states such as the $K_{0,2}^*(1430)$ are essentially outside the low recoil region with $q^2 \gtrsim (14-15) \, \mbox{GeV}^2$, there is some overlap with the $K^*$-region due to their width.
Since the $\phi$ is much more narrow,  backgrounds to $\bar B_s  \to \phi (\to \bar K K)  \ell \ell$ are generically smaller and in particular there is no low lying scalar resonance decaying to $K \bar K$ with an appreciable branching fraction.
 
The plan of the paper is as follows: In Section \ref{sec:Heff} we introduce the effective weak
 $|\Delta B|=|\Delta S|=1$ Hamiltonian used in this work. In Section \ref{sec:matrix} we present the
 $\bar B \to \bar K \pi \ell \ell$ matrix element relevant at low hadronic recoil. Angular distributions are given in Section \ref{sec:angular}. The phenomenology is worked out in Section \ref{sec:pheno}, before we conclude in Section \ref{sec:conclusions}. Auxiliary information on parametric input, kinematics, phase space and
 form factors is deferred to several appendices.

\section{Effective Hamiltonian \label{sec:Heff}}

We employ the following effective hamiltonian for rare $|\Delta B|=|\Delta S|=1$ decays:
\begin{align}
  \label{eq:Heff}
  {\cal{H}}_{\rm eff}= 
   - \frac{4\, G_F}{\sqrt{2}}  V_{tb}^{} V_{ts}^\ast \,\frac{\alpha_e}{4 \pi}\,
     \sum_i C_i(\mu)  {\cal{O}}_i(\mu) \, ,
\end{align}
where 
\begin{align} \nonumber
  {\cal{O}}_{7} & = \frac{m_b}{e} \bar{s} \sigma^{\mu\nu} P_{R} b  F_{\mu\nu}\,, \quad 
   &{\cal{O}}_{7}^\prime &= \frac{m_b}{e} \bar{s} \sigma^{\mu\nu} P_{L} b  F_{\mu\nu}\,,
\\ \label{eq:7910}
  {\cal{O}}_{9} & =  \bar{s} \gamma_\mu P_{L} b \, \bar{\ell} \gamma^\mu \ell \,, \quad
   &{\cal{O}}_{9}^\prime  &=  \bar{s} \gamma_\mu P_{R} b \, \bar{\ell} \gamma^\mu \ell \,,
\\
  {\cal{O}}_{10} & = \bar{s} \gamma_\mu P_{L} b \, \bar{\ell} \gamma^\mu \gamma_5 \ell \,,
  \quad  &{\cal{O}}_{10}^\prime  &=  \bar{s} \gamma_\mu P_R b \, \bar{\ell} \gamma^\mu \gamma_5\ell \,.
\nonumber
\end{align}
Here $P_{L/R}=(1 \mp \gamma_5)/2$ denote chiral projectors, $ \alpha_e$ the fine structure constant, $\mu$ the renormalization scale and $F_{\mu \nu}$ the electromagnetic field strength tensor. Doubly Cabibbo-suppressed
contributions proportional to $V_{ub} V_{us}^*$ are neglected. 

In the  SM the $ {\cal{O}}_{7,9,10}$  induce the dominant contributions, whereas the Wilson coefficients of the chirality-flipped operators, $ {\cal{O}}_{7,9,10}^\prime$, can be of importance in extensions of the SM. We neglect lepton masses in this work.
While we do not specifically study
CP-asymmetries, our framework covers the possibility of  complex Wilson coefficients.
CP-violation at the level of the SM can be taken into account in a straightforward manner.
The operators in Eq.~(\ref{eq:7910}) may be amended by scalar and tensor ones to achieve a complete basis of dimension 6 operators which is beyond the scope of this work.

Lepton-universality breaking effects between $\ell=e$ and $\ell=\mu$ operators can be taken into account as well, assigning  a lepton flavor index to each operator and respective Wilson coefficient 
$C^{ (\prime) \ell}$. Such effects could be probed  for with ratios 
 (with the same cuts in both numerator and denominator, to minimize theory uncertainties)
\begin{align} \label{eq:RFpi}
R_{\bar K \pi} \equiv \frac{ {\cal{B}}(\bar B \to \bar K \pi \mu \mu) }{{\cal{B}}(\bar B \to \bar K \pi e e) }  \, ,
\end{align}
proposed previously for $\bar B \to H \ell \ell$, $H=\bar K^{(*)},  X_s$ decays \cite{Hiller:2003js}, 
or by using further ratios of angular observables. Deviations of $R_{\bar K \pi}$ or $R_H$ in general from unity  can be assigned, up to small kinematic
corrections, to  lepton-universality-breaking new physics, {\it e.g.,} in lepto-quark models or supersymmetric ones with R-parity violation \cite{Bobeth:2007dw}, or 
specific $Z^\prime$ models
\cite{Fox:2011qd,Altmannshofer:2014cfa}.
We emphasize that contributions from intermediate charmonium resonances subsequently decaying to electrons or muons drop out in $R_{\bar K \pi}$ and related ratios, after correcting for differences of radiative tails.
Here and in the following we suppress for brevity the analogous expressions for the \BsKKll{} decay, unless stated otherwise.

\section{The $\boldsymbol{\bar B \to \bar K \pi \ell \ell}$ Matrix Element at Low Recoil \label{sec:matrix}}

The non-resonant decays $\bar B \to \bar K \pi \ell \ell$ are accessible at low hadronic recoil with the
OPE in $1/m_b$. In the basis used in this work, given in Eq.~(\ref{eq:7910}),  the generalized transversity amplitudes can be written by means of universality~\cite{Hiller:2013cza} to lowest order in $1/m_b$ as
\begin{align} \label{eq:trans}
H_{0,\parallel}^{L/R} = C_{-}^{L/R}(q^2) \, 
F_{0,\parallel} (q^2,p^2,\cos \theta_K)\;, \quad H_{\perp}^{L/R} = C_+^{L/R}(q^2) \,  F_\perp (q^2,p^2,\cos \theta_K)\;,  
\end{align}
where $F_i$ ($i=0, \perp, \parallel$) denote the transversity form factors
\begin{align} \nonumber 
F_0 &= \frac{{\cal N}_{nr}}{2}  \bigg[  \lambda^{1/2 }w_+(q^2,p^2,\cos \theta_K)+\frac{1}{p^2}\{(m_K^2-m_\pi^2)\lambda^{1/2}-(m_B^2-q^2-p^2) \lambda^{1/2}_{p}\cos \theta_K\} w_-(q^2,p^2,\cos \theta_K) \bigg]\,,\\
\label{eq:Fi}
F_\parallel &= {\cal N}_{nr}  \sqrt{ \lambda_p \frac{q^2}{p^2}} \, w_-(q^2,p^2,\cos \theta_K)\,, \qquad
F_\perp = \frac{{\cal N}_{nr}}{2}\sqrt{ \lambda \lambda_p \frac{  q^2}{p^2}} \, h(q^2,p^2,\cos \theta_K)\,.
\end{align} 
These are later generalized to include the contributions from the resonances decaying to $\bar K\pi$ final states.
The normalization factor for the non-resonant decays reads
\begin{align} \label{eq:Nnr}
{\mathcal N}_{nr}=\frac{G_F |V_{tb}V_{ts}^*| \alpha_e}{ 2^{7} \pi^4  m_B} \sqrt{ \pi \frac{  \sqrt{\lambda  \lambda_p}}{m_B p^2}} \, .
\end{align}
The  $\bar B \to \bar K \pi $ form factors $w_\pm,h$ are defined in Section \ref{sec:FF}.
We denote the invariant mass squared of the dilepton- and $\bar K\pi$-system by $q^2$ and $p^2$, respectively, whereas $\theta_K$ is the angle between the kaon and the $\bar B$ in the $(\bar K\pi)$ center-of-mass system, where details on the kinematics are given in Appendix~\ref{app:kinematics}. Furthermore, the K\"all\'en function $\lambda(a,b,c)$ is given by $\lambda(a,b,c)=a^2+b^2+c^2-2(ab+ac+cb)$, with the short-hand notations $\lambda =\lambda(m_B^2,q^2,p^2)$
and $\lambda_p =\lambda(p^2,m_K^2,m_\pi^2)$, where $m_M$ and $p_M$ denote the mass and 4-momentum of the meson
$M=B,K,\pi$, respectively.

The transversity amplitudes may be expanded in terms of associated Legendre polynomials $P^m_\ell$ 
\begin{align} \nonumber 
F_0 = \sum_{\ell=0} a_0^\ell(q^2,p^2) \,  P^{m=0}_\ell(\cos \theta_K) \, ,\\
F_\parallel=\sum_{\ell=1} a_\parallel^\ell(q^2,p^2) \, \frac{P^{m=1}_\ell(\cos \theta_K)}{\sin \theta_K} \, ,\label{eq:expand} \\
F_\perp=\sum_{\ell=1} a_\perp^\ell (q^2,p^2)\, \frac{P^{m=1}_\ell(\cos \theta_K)}{\sin \theta_K} \, .\nonumber 
\end{align}
Useful relations to  the Legendre polynomials $P_\ell$ are
 $P_\ell^0 =P_\ell$ and $P^{m=1}_\ell= \sin \theta_K \,  \[ d P_\ell/d \cos \theta_K \]$.  As detailed later in Section~\ref{sec:angular},
we take the dependence on $\sin \theta_K$ and $\phi$ in $F_{\parallel, \perp}$ out of the form factor and  assign it to the angular distribution.
The dilepton system can only carry helicities $0,\, \pm 1$, which by helicity conservation gives the respective range for the
$\bar K \pi$ system. The well-known result from $K_{l4}$ decays~\cite{Kl4} is recovered that there is no S-wave contribution for
$H_{\parallel, \perp}$ since $P^1_0=0$. The $a_i^\ell$ coefficients can be obtained using the orthogonality of the 
$P_\ell^m$, the magnitude of which drops rapidly with increasing $\ell$.

The $q^2$-dependent short-distance coefficients $C_\pm^{L/R}$ are defined as
\begin{align}
 C_\pm^L(q^2)& = C_9^{\rm eff}(q^2) \pm  C_9^\prime - (C_{10}  \pm  C_{10}^\prime)  + \kappa  \frac{2 m_b m_B}{q^2  }
(C_7^{\rm eff} \pm  C_7^\prime) \, , \\
C_\pm^R(q^2)&=C_9^{\rm eff}(q^2) \pm  C_9^\prime+C_{10}  \pm  C_{10}^\prime + \kappa  \frac{2 m_b m_B}{q^2  }
(C_7^{\rm eff} \pm  C_7^\prime)  \, ,
\end{align}
and resemble the ones of  Ref.~\cite{Bobeth:2012vn}
for $B \to K^{(*)} \ell^+ \ell^-$ decays to which we refer for details. The $C_i^{ (\rm eff)}, C_i^\prime$ are (effective) coefficients of the operators in Eq.~(\ref{eq:7910}).
Time-like polarization does not contribute in the limit of vanishing lepton masses.
Corrections to the heavy-quark limit
are parametrically suppressed  as
${\cal{O}}(\alpha_s
\Lambda/m_b, C_7^{(\prime)}/C_9
\Lambda/m_b)$ and at the percent level. Non-factorizable corrections vanish at the kinematic endpoint, {\it i.e.} at zero recoil.

In Section \ref{sec:sd} we give the requisite short-distance couplings in $\bar B \to \bar K \pi  \ell^+ \ell^-$  decays at low recoil. In Section~\ref{sec:FF} we discuss the $\bar B \to \bar K \pi $ form factors
and compute the improved Isgur-Wise relations. In Section~\ref{sec:ep} we present exact relations
for the non-resonant transversity amplitudes
which hold at zero recoil.
For brevity, in the following we frequently suppress the arguments in various phase-space-dependent functions.

\subsection{Short-distance couplings \label{sec:sd}}
In terms of a model-independent analysis the short-distance couplings in Eq.~(\ref{eq:trans}) constitute four complex functions of the dilepton mass. In $\bar B \to \bar K \pi  \ell^+ \ell^-$  decays, the following combinations of these couplings appear:
\begin{align} \label{eq:SD}
\rho_1^\pm =\frac{1}{2} \left(|C_\pm^R|^2 +|C_\pm^L|^2\right) \, , \quad \delta \rho=\frac{1}{4} \left(|C_-^R|^2 -|C_-^L|^2\right) \, , \quad 
\rho_2^\pm  =\frac{1}{4} \left(C_+^R C_-^{R*} \mp C_-^L C_+^{L*}\right) \, ,
\end{align}
where $\rho_2^\pm $ can in general be complex and $\rho_2^+$ equals $\rho_2$ as 
in \cite{Bobeth:2012vn}. 
The corresponding expressions in terms of the (effective) Wilson coefficients read
\begin{align}
\rho_1^\pm & = \left| C_9^{\rm eff} \pm  C_9^\prime + \kappa  \frac{2 m_b m_B}{q^2  }
(C_7^{\rm eff} \pm  C_7^\prime)\right|^2 + |C_{10}  \pm  C_{10}^\prime|^2 \, , \\
\delta \rho & = {\rm Re}\left[ \left(C_9^{\rm eff} -  C_9^\prime + \kappa  \frac{2 m_b m_B}{q^2  } 
(C_7^{\rm eff} -  C_7^\prime) \right)\left(C_{10}  -  C_{10}^\prime\right)^* \right]  \, , \\
{\rm Re} \rho_2^+ & ={\rm Re} \left[\left( C_9^{\rm eff} + \kappa  \frac{2 m_b m_B}{q^2  }
C_7^{\rm eff} \right)C_{10}^*  - \left(C_9^\prime + \kappa  \frac{2 m_b m_B}{q^2  }  C_7^\prime\right)C_{10}^{\prime *} \right]  \, , \\
{\rm Im} \rho_2^+ & = {\rm Im}\left[ C_{10}^{\prime} C_{10}^* + 
\left(C_{9}^{\prime}+\kappa  \frac{2 m_b m_B}{q^2} C_7^{\prime}\right)\left(C_9^{\rm eff} + \kappa  \frac{2 m_b m_B}{q^2  } C_7^{\rm eff}\right)^{\! *}
\right]  \, , \\
{\rm Re} \rho_2^- & = \frac{1}{2} \left[  |C_{10}|^2  -| C_{10}^\prime|^2+ \left|C_{9}^{\rm eff}+\kappa  \frac{2 m_b m_B}{q^2  }C_{7}^{\rm eff}\right|^2  -\left| C_{9}^\prime+\kappa  \frac{2 m_b m_B}{q^2  }C_7^{\prime}\right|^2 
\right]  
\, , \\
{\rm Im} \rho_2^- & = {\rm Im} \left[C_{10}^\prime  \left(C_9^{\rm eff}  + \kappa  \frac{2 m_b m_B}{q^2  } 
C_7^{\rm eff} \right)^* - C_{10}  \left(C_9^\prime + \kappa  \frac{2 m_b m_B}{q^2  }  C_7^\prime\right)^* \right]  \, .
\end{align}
The coefficient $\kappa =  1- 2 \alpha_s/(3 \pi)   \ln(\mu/m_b) $, as in Ref.~\cite{Bobeth:2010wg},
stems from relating the tensor to the vector form factors, as shown below.

The accessibility of the coefficients $\rho_2^-$ and $\delta \rho$ is  
a new feature of the non-resonant decays with respect to $\bar B \to  
\bar K^{(*)} \ell \ell$ decays.
In the SM basis of operators, where the $C_i^\prime$ are negligible, only two couplings exist, $\rho_{1,2}$ \cite{Bobeth:2010wg}, and the following relations hold:
\begin{align} \label{eq:SMrho}
\rho_1 \equiv \rho_1^\pm=2 {\rm Re} \rho_2^- \, , \quad \rho_2 \equiv {\rm Re} \rho_2^+ =\delta \rho \, , \quad
{\rm Im} \rho_2^\pm=0\, . \hspace{1cm} (\mbox{SM basis})
\end{align}

\subsection{ $\boldsymbol{\bar B \to \bar K \pi}$ form factors \label{sec:FF}}

The relevant $\bar B \to \bar K \pi$ matrix elements can be parameterized as follows:
\begin{align}
\langle\bar K^i(p_K)\pi^j(p_\pi)|\bar s\gamma_\mu(1-\gamma_5)b|
\bar B(p_B)\rangle& 
		=i c_{ij}\left[w_+p_\mu+w_-P_\mu+rq_\mu+ih\epsilon_{\mu\alpha\beta\gamma}p_B^\alpha p^\beta P^\gamma\right]\,, \label{eq::FFLLW} \\
       \langle\bar K^i(p_K)\pi^j(p_\pi)|
\bar s i q^\nu \sigma_{\mu \nu}(1+\gamma_5)   b|\bar B(p_B)\rangle &=
- i c_{ij} m_B \left[ w_+' p_{ \mu} + w_-' P_{\mu} + r' q_{\mu}
       + i h' \varepsilon_{\mu\alpha\beta\gamma}p_B^\alpha p^\beta
       P^\gamma \right]  \, , \label{eq::FFLLW2}
\end{align}
where the form factors $w_\pm^{\scriptscriptstyle{(\prime)}}, r^{\scriptscriptstyle{(\prime)}}, h^{\scriptscriptstyle{(\prime)}}$ depend on $q^2,\,p^2$ and $\cos \theta_K$.
The combinatorial factors are given as $|c_{-+}|^2= |c_{0-}|^2= 2|c_{00}|^2= 2|c_{-0}|^2=1$.
Employing $\epsilon_{0123}=-1$ and $\sigma_{\mu \nu}=i/2 [\gamma_\mu, \gamma_\nu]$, the relations to the form factors adapted from Ref.~\cite{Buchalla:1998mt} are $a=w_+-w_--r$, $b=w_++w_--r$, $c=r$, where details are given in Appendix~\ref{app:nonresonant}.
The form factor $r$, which parameterizes the $q_\mu$ component, does not contribute
in the approximations employed in this work, {\it i.e.}  vanishing lepton masses and absence of scalar operators.
For the (pseudo-)scalar matrix element follows (we neglect the strange quark mass):
\begin{align}
\langle\bar K^i(p_K)\pi^j(p_\pi)|\bar s (1+\gamma_5)b|
\bar B(p_B)\rangle& =
i c_{ij} /m_b\left[ w_+ p \cdot q + w_- P \cdot q + r q^2 \right]  \, .
\end{align}

We compute the improved  Isgur-Wise relations  \cite{Grinstein:2002cz}  to lowest order in $1/m_b$ and including
${\cal{O}}(\alpha_s)$ corrections using the method of equations of motion: starting from
 \begin{align} 
   i \partial^\nu (\bar s i \sigma_{\mu \nu} (1+\gamma_5) b)   =
   - m_b \bar s \gamma_\mu (1-\gamma_5) b +i \partial_\mu (\bar s (1+\gamma_5) b) 
   - 2 \bar s i \!\stackrel{\leftarrow}{D}_{\mu} (1+\gamma_5)  b \, 
\end{align}
and matching onto the heavy quark expansion,  refer to, \emph{e.g.}, Ref.~\cite{Grinstein:2004vb} for the Wilson coefficients, 
we obtain 
\begin{align}  \label{eq:iwr}
w_\pm^\prime &= w_\pm \kappa \, , \quad h^\prime = h  \kappa \, .
\end{align}
Using Eq.~(\ref{eq:iwr}), the universal behavior \cite{Bobeth:2010wg} of the OPE detailed for $\bar B \to \bar K^* \ell^+ \ell^-$  in \cite{Grinstein:2004vb} becomes manifest,  in the leading order matrix element of non-resonant decays, Eq.~(\ref{eq:trans}).
The explicit  results in HH$\chi$PT \cite{Buchalla:1998mt} are consistent with this when
 keeping leading terms in the expressions for the primed form factors  only, 
 as given in Appendix~\ref{app:nonresonant}. 
 The form factors to lowest order in $1/m_b$ used in this paper are given as
\begin{align} \nonumber
w_\pm &= \pm \frac{gf_B}{2 f^2} \frac{m_B}{v\cdot p_\pi+\Delta} \, ,\\
h &= \frac{g^2 f_B}{2f^2} \frac{1}{ 
[v\cdot p_\pi+\Delta][v\cdot p+\Delta+\mu_s]} \, , \label{eq:ffinput}
\end{align}
where $v=p_B/m_B$,
$\Delta=m_{B^*}-m_B=46$ MeV and $\mu_s= m_{B_s}-m_B=87.3$ MeV \cite{PDG}.
Here, $f_{(B)}$ denote the decay constants in the $SU(3)$ limit  of the light and heavy meson multiplets and $g$ the HH$\chi$PT coupling, where the values used are given in Appendix~\ref{app:input}. As is common practice for the phase space as well, we take into account the effect of different meson masses.

The corresponding expressions for the $\bar B_s \to \bar K K$ form factors read:
\begin{align} \nonumber
w_\pm &= \pm \frac{gf_{B}}{2 f^2} \frac{m_{B_s}}{v\cdot p_K+\Delta_s} \, ,\\
h &= \frac{g^2 f_{B}}{2f^2} \frac{1}{ 
[v\cdot p_K+\Delta_s][v\cdot p+\Delta]} \, , \label{eq:ffinputBs}
\end{align}
where $\Delta_s=m_{B^*}-m_{B_s}=-42$ MeV. In this case all combinatorial factors can be set to unity.

HH$\chi$PT is an effective theory that applies to light mesons with soft momenta, sufficiently below the scale of chiral symmetry breaking around 1 GeV.
By kinematical considerations, in $\bar B \to \bar K \pi \ell \ell$ decays the momenta of the final pseudoscalars in the $B$-restframe are smaller for larger
$q^2$-values.
Quantitatively, typical momentum-like scales $E_\pi-m_\pi$ and $E_K-m_K$ \cite{BD92} do not exceed
 0.8 GeV (0.5 GeV) for $q^2$ above 14 (16) $\mbox{GeV}^2$, but are smaller in most of the corresponding $(p^2, \cos\theta_K)$-parameter space. 
  While higher-order corrections in the regions with  larger momenta
 will be more important, the expansion is trustworthy for most of the low-recoil phase space. We employ (\ref{eq:ffinput}) and (\ref{eq:ffinputBs})  for the full low recoil region.

\subsection{Endpoint relations \label{sec:ep}}
  
The transversity amplitudes $H_i^{L/R}$, where $i=0,\parallel, \perp$, of the weak decays 
$\bar B \to \bar K^*_J \ell \ell$, where $\bar K_J^*$ denotes a kaonic meson with spin $J$ and mass $m_{K_J}$, are subject to endpoint relations, that is,
 kinematic constraints at vanishing recoil $\lambda(m_B^2,q^2,m_{K_J}^2)=0$  \cite{Hiller:2013cza}.
This situation corresponds to vanishing 3-momenta of the final hadronic  and leptonic system 
$\vec p=\vec q=0$ in the center-of-mass system of the $B$, leading to an enhanced rotational symmetry because of absence of direction.
The endpoint relations read,~\cite{Hiller:2013cza},\footnote{Relative signs depend on conventions for polarization vectors.}
\begin{align}\nonumber
H_0^{L/R}&=0 +{\cal{O}}\left(\sqrt{\lambda}\right)\, , \quad   &(J=0) \\
H_\perp^{L/R}&=0+{\cal{O}}\left(\sqrt{\lambda}\right) \, , \qquad H_\parallel^{L/R} =-\sqrt{2} H_0^{L/R}+{\cal{O}}(\lambda) \, , \quad   &(J=1)  \label{eq:ep}\\
H_{0,\parallel, \perp  }^{L/R}&=0 +{\cal{O}}\left(\lambda^{(J-1)/2}\right)  \, .\quad   &(J \geq 2) \nonumber
\end{align}
Corrections in the vicinity of the endpoint 
are ruled by parity selection
and are indicated above. In the low recoil OPE the relations are equally present in the hadronic form factors.

The endpoint relations for the non-resonant decays are obtained from angular expansion
at  $\lambda_*=\lambda(m_B^2,q^2,p^2)=0$.
This  corresponds to $q^2_*=(m_B-\sqrt{p^2})^2$ for fixed $p^2$ or
$p^2_*=(m_B-\sqrt{q^2})^2$ for fixed $q^2$. In particular, $\lambda_p$ is finite at $p_*^2$ in general.
Eq.~(\ref{eq:ep})  implies that the endpoint is dominated by the $\ell=1$ amplitudes with
longitudinal and parallel polarization which are related and finite after removing the common phase space factor ${\cal{N}}_{nr}$.
To show this explicitly in the low recoil OPE, define $\hat F_i \equiv F_i/{\cal{N}}_{nr}$ and denote the corresponding coefficients as in Eq.~(\ref{eq:expand}) by $\hat a_i^\ell$. At the endpoint one readily obtains that
all  $\hat a_i^\ell$ vanish except for $\hat a_{0,\parallel}^1$, which obey
\begin{align} \label{eq:a-relation}
\hat a_{0}^1=\hat a_{\parallel}^1=\left.-\sqrt{ \frac{q^2}{p^2} \lambda_p}\,\, w_- \right|_{\lambda=\lambda_*} \, ,
\end{align}
which is consistent with Eq.~(\ref{eq:ep}). Note that due to isotropicity, the form factor $w_-$ is $\cos \theta_K$-independent at the endpoint, in accordance with the HH$\chi$PT results
Eq.~(\ref{eq:ffinput}).
 
It follows that the relations for $\bar B \to \bar K^* \ell \ell$ decay observables \cite{Hiller:2013cza}  hold at $\lambda_*$ for the non-resonant decay, including the fraction of longitudinally polarized $(\bar K \pi)$, $F_L$, being $1/3$. Note that the endpoint values are assumed in general at different values of $q^2$ such that the non-resonant modes do dilute the vector signal predictions. Phenomenological consequences are discussed further in Sections \ref{sec:angular} and \ref{sec:kstar}.

\section{Angular Distributions  \label{sec:angular}}

We present angular distributions for $\bar B \to \bar K \pi \ell \ell$  decays for the basis given in Eq.~(\ref{eq:7910}) and massless leptons.
In Section \ref{sec:general} we give the general expressions and discuss the special point of zero recoil.
In Section \ref{sec:resonant} we discuss angular projections, where the $\bar K \pi$ system is
in low angular momentum configuration, $\ell=0,1,2$, $i.e.$,  S, P, D partial waves.  This expansion corresponds to the lowest order terms in non-resonant decays as well as resonant contributions from spin $J=0,1,2$ states subsequently decaying to  $\bar K \pi$. In Section \ref{sec:OPE} we present the angular distributions in the low recoil region based on Eq.~(\ref{eq:trans}).

\subsection{General case \label{sec:general}}

 The $\bar B \to \bar K \pi \ell \ell$ angular distribution, with the angles $\theta_\ell, \theta_K, \phi$ defined as in \cite{Bobeth:2008ij},
can be written as
  \begin{eqnarray}\label{eq:d5Gamma}
d^5\Gamma 
&=&\frac{1}{ 2  \pi} \left[ \sum c_i(\theta_\ell,\phi) I_i \left(q^2,p^2,\cos \theta_K\right) \right] dq^2dp^2d\cos\theta_Kd\cos\theta_\ell d\phi\,,
\end{eqnarray}
where 
\begin{align}
c_1 & =1, \quad c_2=\cos 2\theta_\ell, \quad c_3=\sin^2\theta_\ell\cos 2\phi, \quad c_4=\sin 2\theta_\ell \cos \phi, \quad c_5=\sin\theta_\ell\cos\phi, \nonumber \\ c_6& =\cos\theta_\ell, \quad c_7=\sin\theta_\ell\sin\phi, \quad c_8=\sin 2\theta_\ell\sin\phi, \quad c_9=\sin^2\theta_\ell\sin2\phi \, .
\label{eq:ci}
\end{align}
The phase space allows the angles to be within the ranges
\begin{align}
-1 < \cos \theta_K \leq 1 \, , \quad -1 < \cos \theta_\ell  \leq 1 \, , \quad  0< \phi  \leq 2 \pi \, .
\end{align}
The coefficient functions $I_i \equiv I_i(q^2,p^2,\cos \theta_K)$  are given in terms of transversity amplitudes in Eq.~(\ref{eq:trans}) as 
\begin{align}  \nonumber
I_1 & = \phantom{-}\frac{1}{16} \bigg[ |\azeL|^2 +|\azeR|^2 + \frac{3}{2}\sin^2 \theta_K \{ |\apeL|^2 + |\apaL|^2 + (L\to R) \} \bigg],
\\  \nonumber
  I_2 & = -\frac{1}{16} \bigg[ |\azeL|^2 + (L\to R)  -\frac{1}{2} \sin^2 \theta_K \{ |\apeL|^2+ |\apaL|^2 + (L\to R) \}\bigg],
\\  \nonumber
  I_3 &=  \phantom{-}\frac{1}{16}   \bigg[ |\apeL|^2 - |\apaL|^2  + (L\to R)\bigg]  \sin^2\theta_K,
\\  \nonumber
  I_4 & = -\frac{1}{8}  \bigg[\re (\azeL^{}\apaL^*) + (L\to R)\bigg] \sin\theta_K,
\\  \label{IH}
  I_5 & = -\frac{1}{4}  \bigg[\re(\azeL^{}\apeL^*) - (L\to R)\bigg] \sin\theta_K,
\\  \nonumber
  I_6 & = \phantom{-}\frac{1}{4}   \bigg[\re (\apaL^{}\apeL^*) - (L\to R)\bigg] \sin^2\theta_K,
\\  \nonumber
  I_7 &=- \frac{1}{4} \bigg[\im (\azeL^{}\apaL^*) - (L\to R)\bigg] \sin\theta_K,
\\  \nonumber
  I_8 & =-\frac{1}{8}  \bigg[\im(\azeL^{}\apeL^*) + (L\to R)\bigg] \sin\theta_K,
\\  \nonumber
  I_9 & =\phantom{-}\frac{1}{8}   \bigg[\im (\apaL^{*} \apeL) + (L\to R)\bigg] \sin^2\theta_K .
\end{align}

After integrating over $\phi, \cos \theta_\ell$ and both, respectively, we obtain
  \begin{eqnarray}
\frac{d^4\Gamma }{dq^2dp^2d\cos\theta_Kd\cos\theta_\ell}
&=&  I_1  +I_2 \cos  2\theta_\ell  +I_6  \cos \theta_\ell   \,, \\
\frac{d^4\Gamma }{dq^2dp^2d\cos\theta_Kd \phi}
&=&\frac{1 }{ \pi } \left(   I_1  - \frac{I_2}{3}   + \frac{\pi}{4} I_5  \cos \phi + \frac{\pi}{4}  I_7 \sin \phi + \frac{2}{3} I_3 \cos 2 \phi +  \frac{2}{3} I_9 \sin 2 \phi  \right) \,, \\
\frac{d^3\Gamma }{dq^2dp^2d\cos \theta_K }
&=&2  \left(   I_1  - \frac{I_2}{3}  \right) \,. \label{eq:dq2p2cth}
\end{eqnarray}

At zero recoil $\lambda=\lambda_*$, see Section \ref{sec:ep}, the following exact relations hold 
\begin{align}
I_3=-\frac{I_1+I_2}{2}\, , \quad 
I_4 = - \sqrt{ \frac{( I_1+I_2)(I_1-3 I_2)}{2} } \, , \quad
I_{5,6,7,8,9} =0 \,,
\end{align}
and in addition the $\cos \theta_K$ and $\cos \theta_\ell$-distributions become isotropic,
\begin{align}
 \left.\frac{ d^3 \Gamma}{  d q^2 d p^2 d \cos \theta_K}  \Big  / \left( \frac{d^2 \Gamma}{d q^2 d p^2} \right)\right|_{\lambda=\lambda_*}&
\!\!\!\! = \frac{1}{2} \, , \\
 \left.\frac{ d^3 \Gamma}{  d q^2 d p^2 d \cos \theta_l}  \Big  / \left( \frac{d^2 \Gamma}{d q^2 d p^2} \right) \right|_{\lambda=\lambda_*}&
\!\!\!\! = \frac{1}{2} \, ,
\end{align}
while the $\phi$-distribution between the $\bar K \pi$ and $\ell \ell$ planes  does not, 
\begin{align}
\left. \frac{ d^3 \Gamma}{  d q^2 d p^2 d \phi}  \Big  / \left( \frac{d^2 \Gamma}{d q^2 d p^2} \right) \right|_{\lambda=\lambda_*}&
 = \frac{1}{2 \pi} \left(1- \frac{1}{3} \cos 2 \phi\right) \, , 
 \end{align}
in agreement with~\cite{Hiller:2013cza}.

\subsection{S-, P- and D-wave angular projections \label{sec:resonant}}

Using Eq.~(\ref{IH}) and substituting in the Legendre polynomials
\begin{align}
P^0_0(x)=1 \, , P^0_1(x)=x \, , P^0_2(x)=\frac{1}{2} (3 x^2 -1)\, ,  P_1^1(x)=-\sqrt{1-x^2} \, , 
P_2^1(x)=-3 x \sqrt{1-x^2} \, ,
\end{align}
we obtain  the full angular distribution $d^5 \Gamma(S+P+D)$ for
the $\bar K \pi$ pair in S-,\,P- and D-wave configuration.
Since $\ell$ is fixed, all angular dependence can be made explicit.
In particular, the angular coefficients $J_{ix}=J_{ix}(q^2,p^2)$,  stemming from $I_i, \, i=1,..,9$ and given in Appendix~\ref{app:resonant}, do not depend on $\theta_K$. Higher partial waves can be  included in a similar manner. The angular distribution is given as
\begin{align} \nonumber
& \frac{d^5  \Gamma (S+P+D)}{dq^2dp^2d\cos\theta_Kd\cos\theta_\ell d \phi} =\frac{1}{2 \pi}
    \bigg[   \sum_{i=1,2} c_i \left(J_{icc } \cos^2\!\theta_K   + J_{iss} \sin^2\!\theta_K   + J_{ic} \cos \!\theta_K  \right. \\ \nonumber
     &\left.+  J_{issc}  \, \sin^2\! \theta_K \cos\! \theta_K + J_{isscc} \,  \sin^2\! \theta_K \cos^2\! \theta_K\right) \\
   &  + \sum _{i=3,6,9} c_i  \left( J_{icc}  \cos^2\!\theta_K   + J_{i}   +J_{ic} \cos \!\theta_K  \right)  \sin^2\!\theta_K  \nonumber \\
  &  +\sum_ {i=4,5,7,8} c_i   \left(J_{icc}  \cos^2\!\theta_K   + J_{iss} \sin^2\!\theta_K   + J_{ic} \cos \!\theta_K +  J_{issc} \sin^2\! \theta_K  \cos\! \theta_K \right) \sin \!\theta_K 
 \bigg]\, .
\label{eq:full}
\end{align}
Explicit expressions of the $J_{ix}$ are given in Appendix \ref{app:resonant}.
Integration over $\theta_K$ yields 
\begin{align} \nonumber
\frac{d^4\Gamma (S+P+D)}{dq^2dp^2 d\cos\theta_\ell d \phi} &=\frac{1}{2 \pi}
    \left[   \sum_{i=1,2} c_i \left(\frac{2}{3}J_{icc }    + \frac{4}{3}J_{iss}   +\frac{4}{15}J_{isscc}  \right)+ \!\!\sum _{i=3,6,9}\!\! c_i \left(\frac{4}{15}J_{icc }    +  \frac{4}{3}J_{i} \right)\right. \\
   &\qquad\quad\left.  
    +\!\!\!\sum_ {i=4,5,7,8}\!\!\!\! c_i \left(\frac{\pi}{8}J_{icc }    + \frac{3 \pi}{8} J_{iss} \right)
 \right]\, ,
\end{align}
and integrating further over $\phi$ and $\theta_\ell$,
\begin{align} 
\frac{d^2\Gamma (S+P+D)}{dq^2dp^2}& =
   \frac{4}{3}   \left[ J_{1cc }    + 2 J_{1ss}   +\frac{2}{5}J_{1sscc}  -\frac{1}{3} \left(J_{2cc }    + 2 J_{2ss}   +\frac{2}{5} J_{2sscc} \right) \right] \, .
\end{align}

For a pure P-wave, only a subset of coefficients contribute: $J_{iss,icc}$ for $i=1,2$, $J_i$ for $i=3,6,9$ and $J_{ic}$ for $i=4,5,7,8$. Their relation to the ones from $\bar B \to \bar K^* (\to \bar K \pi) \ell \ell$ analyses \cite{Bobeth:2010wg}
(BHV) are: $J_{iss}=3/4J^{BHV}_{is},  J_{icc}=3/4J^{BHV}_{ic}$ ($i=1,2$), $J_i=3/4 J_i^{BHV}$ ($i=3,6,9$)
and $J_{ic} =3/2 J_i^{BHV}$ ($i=4,5,7,8$).

To approximate the non-resonant distribution, {\it i.e.}, its form factors Eq.~(\ref{eq:Fi}) by  its S+P+D-wave components turns out to be useful when discussing the P-wave decays $\bar B \to \bar K^* (\to \bar K \pi) \ell \ell$  to which the non-resonant ones constitute a background. 
The S+P+D distribution will receive two types of corrections from higher waves:
there will be additional, higher trigonometric polynomials of the angle $\theta_K$, and secondly
the S+P+D coefficients $J_{ix}$ will receive further contributions.
Quantitatively, we find that near $p^2=m_{K^*}^2$ the S+P+D approximation
in the longitudinal part of the rate  is at the few permille level, whereas the corrections to
the parallel and perpendicular ones are equal and drop from one percent at $q^2=14 \, \mbox{GeV}^2$ to sub-permille towards zero recoil. The corrections in the simpler S+P approximation  are about one order of magnitude larger. In our numerical estimate we used the explicit form factors Eq.~(\ref{eq:ffinput}).

\begin{figure}
\centering{
\includegraphics[height=0.3\textwidth]{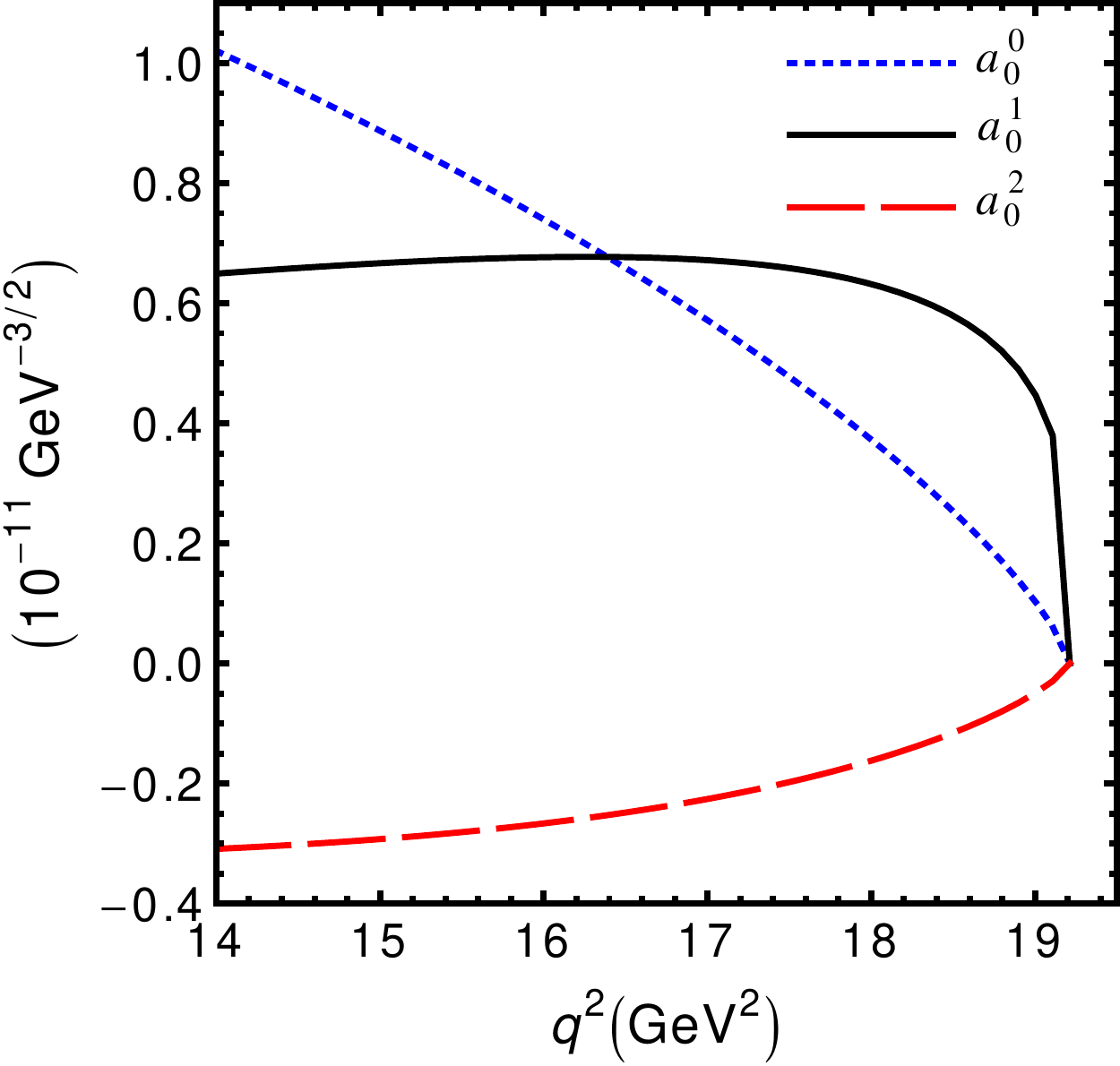}~~
\includegraphics[height=0.3\textwidth]{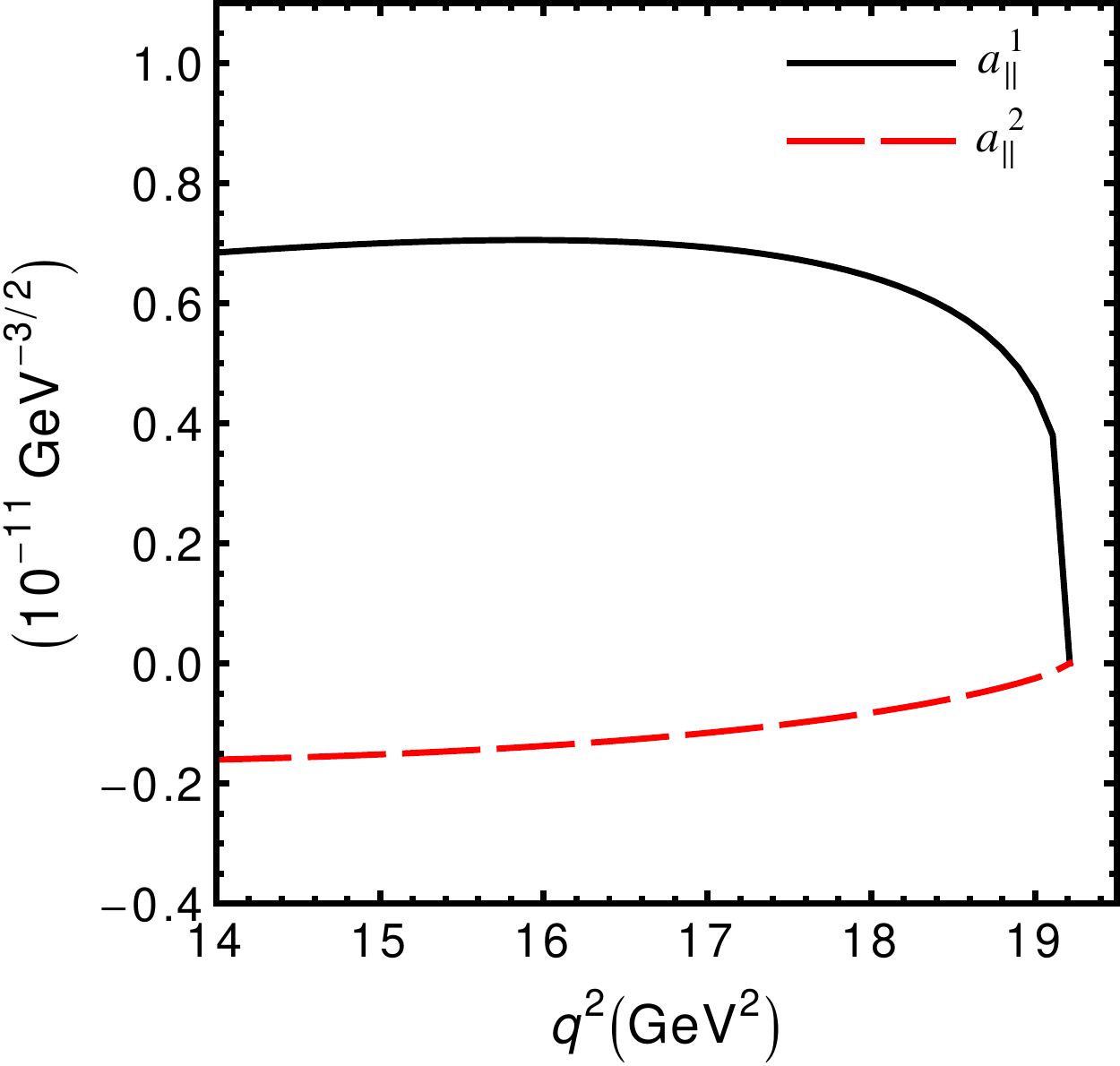}~~
\includegraphics[height=0.3\textwidth]{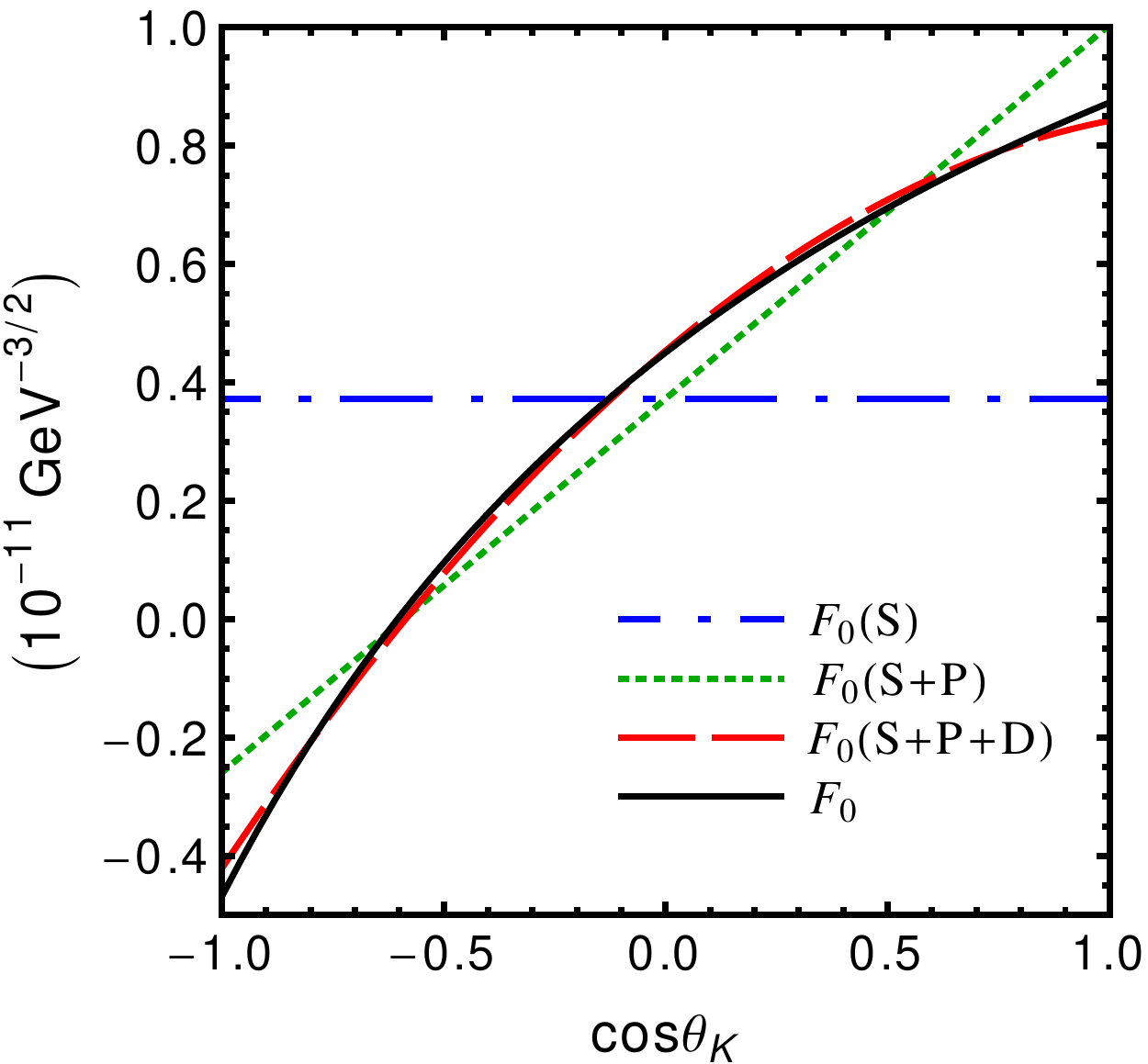}
}
\caption{\small The first few angular coefficients $a_i^\ell$ (left $i=0$, middle $i=\parallel$) of the non-resonant form factors $F_i$, given in Eq.~(\ref{eq:Fi}), for central values of the input parameters at $p^2=m_{K^*}^2$. The blue dotted, black solid and red dashed lines correspond to S, P and D coefficients, respectively.
In the plot to the right the form factor $F_0$ is shown at $p^2=m_{K^*}^2$ and $q^2=18 \, \mbox{GeV}^2$
in S (blue short-long dashed), S+P (green dotted) and S+P+D (red dashed) approximation, together with the full result (solid black curve). }
\label{fig:aSPD}
\end{figure}

To illustrate the features of the S+P+D approximation we show
in Fig.~\ref{fig:aSPD} the first few angular coefficients $a_{0, \parallel}^\ell(q^2,p^2)$  of the non-resonant form factors $F_i$, given in Eq.~(\ref{eq:Fi}), for central values of the input parameters at $p^2=m_{K^*}^2$. Due to the  identical angular dependence of the respective form factors $w_-$ and $h$ the angular expansion for the transverse form factors $F_\parallel$ and $F_\perp$ is identical up to an overall kinematic factor and for brevity $a_\perp^\ell$ is not shown.
Towards lower $q^2$-values the S-wave contribution dominates $F_0$.
Also shown in Fig.~\ref{fig:aSPD}  is $F_0( 18 \, \mbox{GeV}^2, m_{K^*}^2, \cos \theta_K)$ 
in S, S+P and S+P+D approximation, and the full result.
Note that convergence of the angular expansion is achieved after $\cos \theta_K$-integration rather than locally.
Since going to D-waves corresponds to one order more for $F_0$ relative to the  two transverse form factors (not shown) the approximation works better for the former.

The observables based on the $I_i$, and the derived $J_{ix}$, in the $\bar B \to \bar K^* (\to \bar K \pi) \ell \ell$ analyses contain different contributions from the non-resonant S,P and D-wave states.
Generically, the following features hold, where we extend
existing findings for the  S-wave background \cite{Becirevic:2012dp,Matias:2012qz,Blake:2012mb,Bobeth:2012vn} to include D-wave effects:

\begin{itemize}

\item[\it i)] There is no S-wave contribution to $I_{3,6,9}$.

\item[\it ii)] The D-wave contributions to $I_{3,6,9}$ can be separated from the pure P-one by an angular analysis, since both appear individually in separate coefficients $J_{ix}$.

\item[\it iii)] In $I_{1,2}$ the S- and D-wave contributions need to be subtracted by sideband measurements.

\item[\it iv)] The S-P and D-P interference to $I_{4,5,7,8}$ can be separated from the pure P-wave contribution by angular analysis, while S-D and pure D-contributions remain. In general, the latter require sideband subtractions, unless they can be neglected, see below.

\item[\it v)] In the presence of a sufficiently dominating P-wave contribution, the S- and D-waves can be considered small and S-S, S-D and D-D contributions are doubly suppressed and hence may be neglected. This concerns about half of the 35 coefficients $J_{ix}$, given in Appendix~\ref{app:resonant}, which receive in this approximation P-wave contributions, only.

\item[\it vi)] As discussed further in Section~\ref{sec:angularpheno}, strong phase differences are experimentally accessible and signal the presence of interference.

\item[\it vii)] The separation of non-resonant contributions from resonant ones in the same partial wave 
requires sideband subtractions.

\end{itemize}

 \subsection{Low recoil \label{sec:OPE}}

 At leading order low recoil OPE, the sensitivity of the angular coefficients to long- and short-distance physics factorizes 
as follows:
\begin{align} \nonumber
I_1 & = \phantom{-}\frac{1}{8} \left[   |{\cal F}_0 |^2 \rho_1^- +\frac{3}{2} \sin^2 \theta_K  \{  |{\cal F}_\parallel |^2 \rho_1^-+  |{\cal F}_\perp |^2 \rho_1^+\} \right]  \, ,  \\ \nonumber
I_2 & =- \frac{1}{8} \left[   |{\cal F}_0 |^2 \rho_1^- -\frac{1}{2} \sin^2 \theta_K  \{  |{\cal F}_\parallel |^2 \rho_1^-+  |{\cal F}_\perp |^2 \rho_1^+\} \right]   \, ,  \\ \nonumber
I_3 & = \phantom{-}\frac{1}{8}    \left[ |{\cal F}_\perp |^2 \rho_1^+ -  |{\cal F}_\parallel |^2 \rho_1^-  \right]  \sin^2 \theta_K  \, ,  \\ \nonumber
I_4 &= - \frac{1}{4} {\rm Re}({\cal F}_0 {\cal F}_\parallel^*) \,\rho_1^-  \sin \theta_K  \, ,  \\
\label{eq:Iope}
I_5 &=  \phantom{-}\left[{\rm Re}({\cal F}_0 {\cal F}_\perp^*)  {\rm Re} \rho_2^++ {\rm Im} ({\cal F}_0 {\cal F}_\perp^*) 
{\rm Im} \rho_2^-  \right] \sin \theta_K  \, ,  \\ \nonumber
I_6&= - \left[{\rm Re}({\cal F}_\parallel {\cal F}_\perp^*) {\rm Re} \rho_2^+ + {\rm Im} ({\cal F}_\parallel {\cal F}_\perp^*) {\rm Im} \rho_2^-  \right]\sin^2 \theta_K  \, ,   \\ \nonumber
I_7&=   {\rm Im} ({\cal F}_0 {\cal F}_\parallel^*)\,\delta \rho\,  \sin \theta_K \, ,   \\ \nonumber
I_8&=  \frac{1}{2} \left[ {\rm Re}({\cal F}_0 {\cal F}_\perp^*) {\rm Im} \rho_2^+  -  {\rm Im} ({\cal F}_0 {\cal F}_\perp^*) {\rm Re} \rho_2^-  \right] \sin \theta_K \, ,   \\ \nonumber
I_9&=  \frac{1}{2} \left[{\rm Re}({\cal F}_\perp {\cal F}_\parallel^*) {\rm Im} \rho_2^+ + {\rm Im} ({\cal F}_\perp {\cal F}_\parallel^*) {\rm Re} \rho_2^- \right]\sin^2 \theta_K   \, ,  \nonumber
\end{align}
where the short-distance coefficients are given in Eq.~(\ref{eq:SD}) and the generalized
transversity form factors are defined as
\begin{align} \label{eq:fullformfactor}
{\cal F}_0 & \equiv {\cal F}_0\left(q^2, p^2, \cos \theta_K\right)=F_0\left(q^2, p^2, \cos \theta_K\right)+\sum_R P^{0}_{J_R} (\cos\theta_K) \cdot F_{0 J_R}\left(q^2, p^2\right)  \, ,   \\
{\cal F}_i & \equiv  {\cal F}_i\left(q^2, p^2, \cos \theta_K\right)=F_i\left(q^2, p^2, \cos \theta_K\right)+\sum_R \frac{P^{1}_{J_R} (\cos\theta_K) }{\sin \theta_K}  \cdot F_{iJ_R}\left(q^2, p^2\right)  \, , \quad i= \parallel, \perp \, . \nonumber
\end{align}
Here, the  first terms on the right-hand sides are the non-resonant form factors, as given in Eq.~(\ref{eq:Fi}), and the second terms denote contributions from resonances $R$ with spin $J_R$ decaying to $\bar K \pi$  with the corresponding polarization-dependent form factors $F_{(0, \parallel, \perp) J_R}$. The latter can include either a parameterization of the line shape or, in the narrow-width approximation, a delta distribution.
The separation of the individual
contributions to the partial waves is non-trivial, in particular for very wide resonances such as the $\kappa(800)$.
This means that there is a risk to double-count contributions when the line shapes are extracted experimentally.

The factorization of long- and short-distance factors with universal short-distance coefficients at low recoil can be seen in Eq.~(\ref{eq:Iope}). This separation allows suitable observables to be formed that are sensitive to the electroweak physics, 
without the need for separating each of the different contributions to Eq.~(\ref{eq:fullformfactor}).
Strong phase differences between the generalized form factors ${\cal F}_i$ can arise
from the interference of non-resonant decays with resonances or overlapping resonances. This provides an opportunity to probe the couplings $\rho_2^-$ and $\delta \rho$, which otherwise could not be accessed in $\bar B\to \bar K^{(*)} \ell\ell$ decays.\footnote{These couplings can also be accessed in baryonic decays such as $\Lambda_b\to\Lambda\ell\ell$. We thank Danny van Dyk for informing us about their forthcoming publication~\cite{vanDyk2014}.}
At the same time information on strong phases can be extracted experimentally. Particularly useful in this regard are the (naive) T-odd observables  $I_{7,8,9}$ \cite{Bobeth:2008ij}.
 
In the SM basis, Eq.~(\ref{eq:Iope}) can be simplified using Eq.~(\ref{eq:SMrho}) to give
\begin{align} \nonumber
I_1 & = \phantom{-}\frac{1}{8}  \rho_1 \left[   |{\cal F}_0 |^2  +\frac{3}{2} \sin^2 \theta_K  \{  |{\cal F}_\parallel |^2 +  |{\cal F}_\perp |^2 \} \right]  \, , \\ \nonumber
I_2 & =- \frac{1}{8}  \rho_1 \left[   |{\cal F}_0 |^2  -\frac{1}{2} \sin^2 \theta_K  \{  |{\cal F}_\parallel |^2 +  |{\cal F}_\perp |^2 \} \right] \, , \\ \nonumber
I_3 & = \phantom{-}\frac{1}{8}  \rho_1   \left[ |{\cal F}_\perp |^2  -  |{\cal F}_\parallel |^2   \right]   \sin^2 \theta_K \, ,\\ \nonumber
I_4 &= - \frac{1}{4}  \rho_1\, {\rm Re}({\cal F}_0 {\cal F}_\parallel^*)  \sin \theta_K \, , \\
\label{eq:IopeSM}
I_5 &=  \phantom{-}\rho_2\, {\rm Re}({\cal F}_0 {\cal F}_\perp^*)  \sin \theta_K \, , \hspace{4cm} (\mbox{SM basis})\\ \nonumber
I_6&= -   \rho_2 \,{\rm Re}({\cal F}_\parallel {\cal F}_\perp^*) \sin^2 \theta_K \, , \\ \nonumber
I_7&=   \phantom{-}\rho_2 \,{\rm Im}({\cal F}_0 {\cal F}_\parallel^*)  \sin \theta_K \, ,\\ \nonumber
I_8&= - \frac{1}{4} \rho_1\, {\rm Im}({\cal F}_0 {\cal F}_\perp^*)  \sin \theta_K \, ,\\ \nonumber
I_9&=  \phantom{-}\frac{1}{4} \rho_1\, {\rm Im} ({\cal F}_\perp {\cal F}_\parallel^*) \sin^2 \theta_K  \, . \nonumber
\end{align}
Note that the non-vanishing values of $I_{7,8,9}$ in the SM basis are induced by non-vanishing relative strong phases.

\section{Phenomenological analysis \label{sec:pheno}}

A key feature of non-resonant $\bar B \to \bar K \pi \ell \ell$ decays is that at low recoil they are amenable to the lowest order OPE, resulting in Eq.~(\ref{eq:Iope}). As in
$\bar B \to \bar K^{(*)} \ell \ell$ decays \cite{Bobeth:2010wg,Bobeth:2012vn}, the separation of  short-distance from form factor coefficients allows to construct observables that are sensitive to 
short-distance or hadronic physics separately.  This makes non-resonant decays useful by themselves, given sufficient data for an angular analysis. Here we discuss only a few important phenomenological applications, leaving a more detailed analysis for future work.
The numerical
estimates are based on HH$\chi$PT form factors, as given in Eq.~(\ref{eq:ffinput}), which are extrapolated in parts of the phase space beyond their nominal region of validity. This means that the uncertainties in particular in rates covering such regions are possibly underestimated.
This highlights the importance of ratios and asymmetries to be constructed from the angular analysis, which can have a much weaker dependence on form factors, as well as independent and improved $\bar B \to \bar K \pi$ form factor determinations.

We perform phenomenological studies at low hadronic recoil of non-resonant $\bar B \to \bar K \pi \ell \ell$ decays (Section \ref{sec:kstar}), discuss aspects of the angular analysis (Section~\ref{sec:angularpheno}), including resonances (Section \ref{sec:kstarreso}) and of  $\bar B_s  \to \bar K K  \ell \ell$  decays (Section \ref{sec:phi}).

\subsection{Non-resonant $\boldsymbol{\bar B \to \bar K \pi \ell \ell}$ decays  \label{sec:kstar}}
In the analysis of non-resonant \BKpill{} decays, the focus lies on estimating their influence on $\bar B \to \bar K^* ( 892) \ell \ell$  (``P-wave'') analyses. To that aim, the
ranges of interest for the invariant mass of the $(\bar K\pi)$ system are defined as follows:
\begin{itemize}
\item[--] Full phase space of the non-resonant decay:  $p^2_{\rm min} \equiv (m_K+m_\pi)^2 \leq p^2 < (m_B- \sqrt{q^2})^2$, where the endpoint for the dilepton system is 
$q^2=q^2_{\rm max} \equiv (m_B-\sqrt{p^2_{\rm min}})^2=21. 58\, \mbox{GeV}^2$.
\item[--] P-wave 'signal' window: $0.64  \, \mbox{GeV}^2 \leq p^2 < 1 \, \mbox{GeV}^2$,  corresponding to the endpoint $q^2=20.06 \, \mbox{GeV}^2$.
\item[--] S+P-wave 'total' window:   $p^2_{\rm min} \leq p^2 < 1.44 \, \mbox{GeV}^2$, corresponding to the endpoint $q^2=q^2_{\rm max}$. 
\end{itemize}
We stress that  both, the signal $\bar B \to \bar K^* \ell \ell$ and the non-resonant $\bar B\to \bar K\pi\ell\ell$ background decays, are $|\Delta B|=|\Delta S|=1$ FCNCs and
need to be analyzed together in a model-independent way.

\begin{figure}
\centering{
\includegraphics[height=0.4\textwidth]{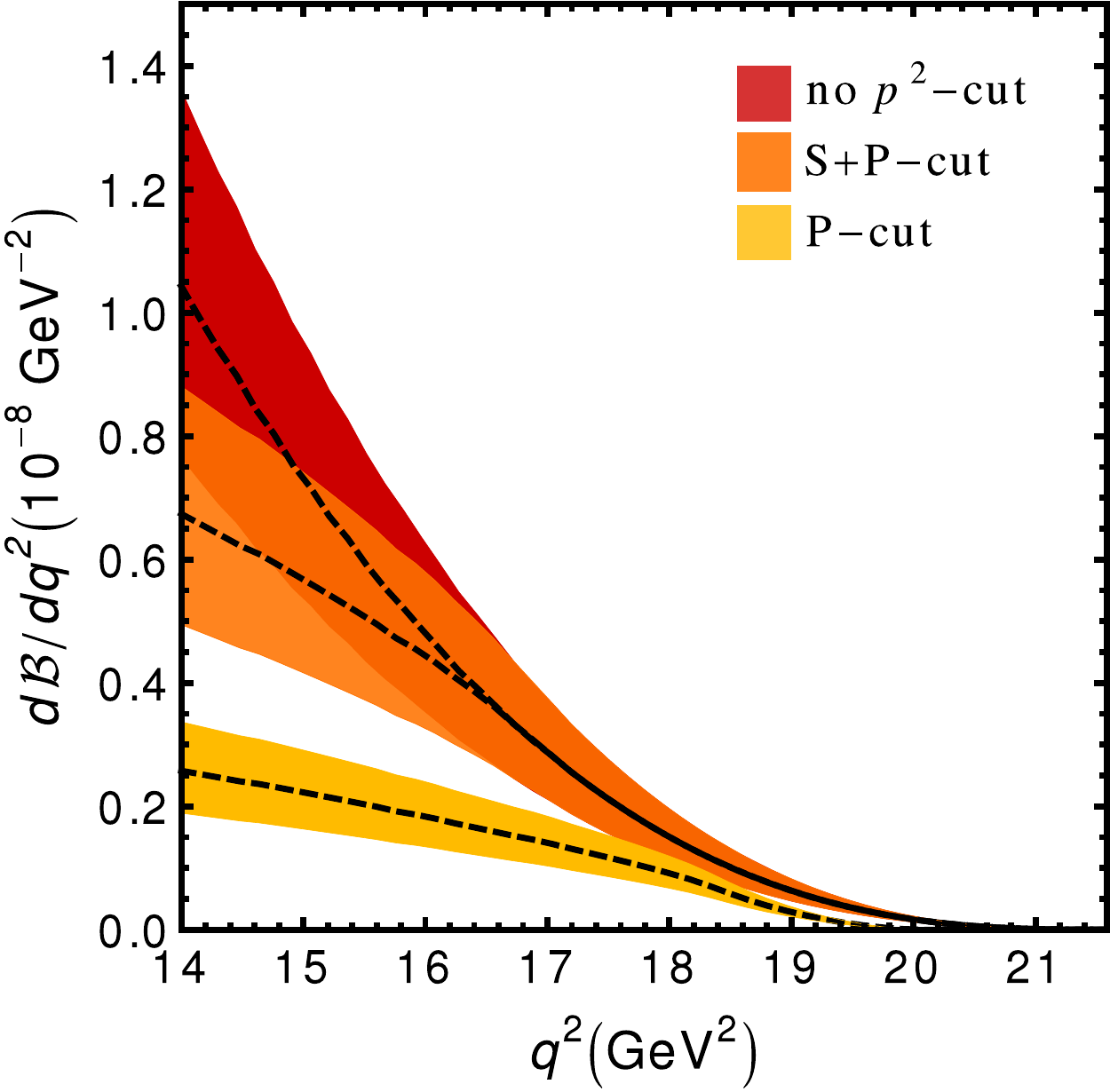}\qquad\quad
\includegraphics[height=0.4\textwidth]{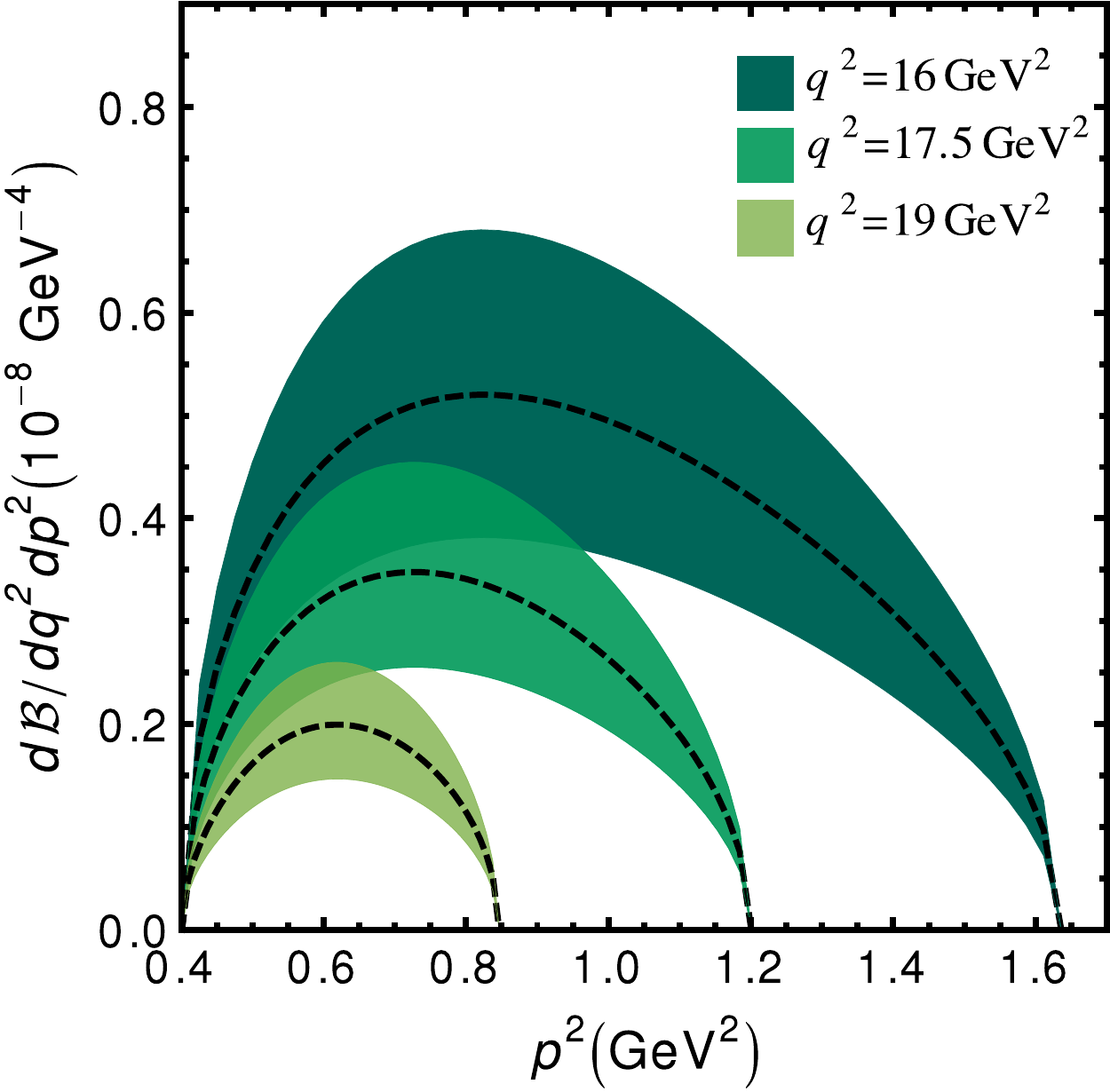}
}
\caption{\small The non-resonant differential branching fraction $d{\cal{B}}(\bar B \to \bar K \pi \ell \ell)/dq^2$ (left)  without $p^2$-cuts, in the $P$-wave 'signal' window and the S+P-wave 'total' window, and 
$d^2{\cal{B}}(\bar B \to \bar K \pi \ell \ell)/d q^2 dp^2$  (right) for fixed $q^2=16,17.5,19 \, \mbox{GeV}^2$ (from outer to inner curves) 
in the SM, see text for details. $\bar B \to \bar K \pi$ form factors are taken from HH$\chi$PT,  Eq.~(\ref{eq:ffinput}), and include parametric uncertainties only. Dashed lines are for central values of the input parameters. }
\label{fig:BKpi}
\end{figure}

Differential SM branching ratios for the non-resonant decays are shown in Fig.~\ref{fig:BKpi}. 
The curves are obtained by integrating Eq.~(\ref{eq:dq2p2cth}) over the accessible phase space.
Form factors from HH$\chi$PT are employed, as given in Eq.~(\ref{eq:ffinput}), using the parametric input given in Table~\ref{tab:input}.  
The left-hand plot of Fig.~\ref{fig:BKpi} shows the impact of the $p^2$-cuts defined above on the $q^2$-distribution.
Without $p^2$ cut, the resulting integrated SM branching fraction in the low-recoil region is in the few $10^{-8}$ range,
about an order of magnitude smaller than the corresponding ones for $\bar B \to \bar K^* \ell \ell$, as given, {\it e.g.,} in \cite{Bobeth:2012vn}. 
Since the very high-$q^2$ region is dominated by small hadronic masses, the distribution for the S+P window differs from the one without cuts only at smaller values of $q^2$. This feature is shared also in the following plots with other observables. Both,
P- and S+P-window cuts reduce the  low-recoil branching ratios to around $10^{-8}$. The right-hand plot of  Fig.~\ref{fig:BKpi} shows the $p^2$ distribution for fixed values of $q^2$. The spectrum is very different from the Breit-Wigner resonance distributions.

\begin{figure}[htb]
\centering{
\includegraphics[height=0.4\textwidth]{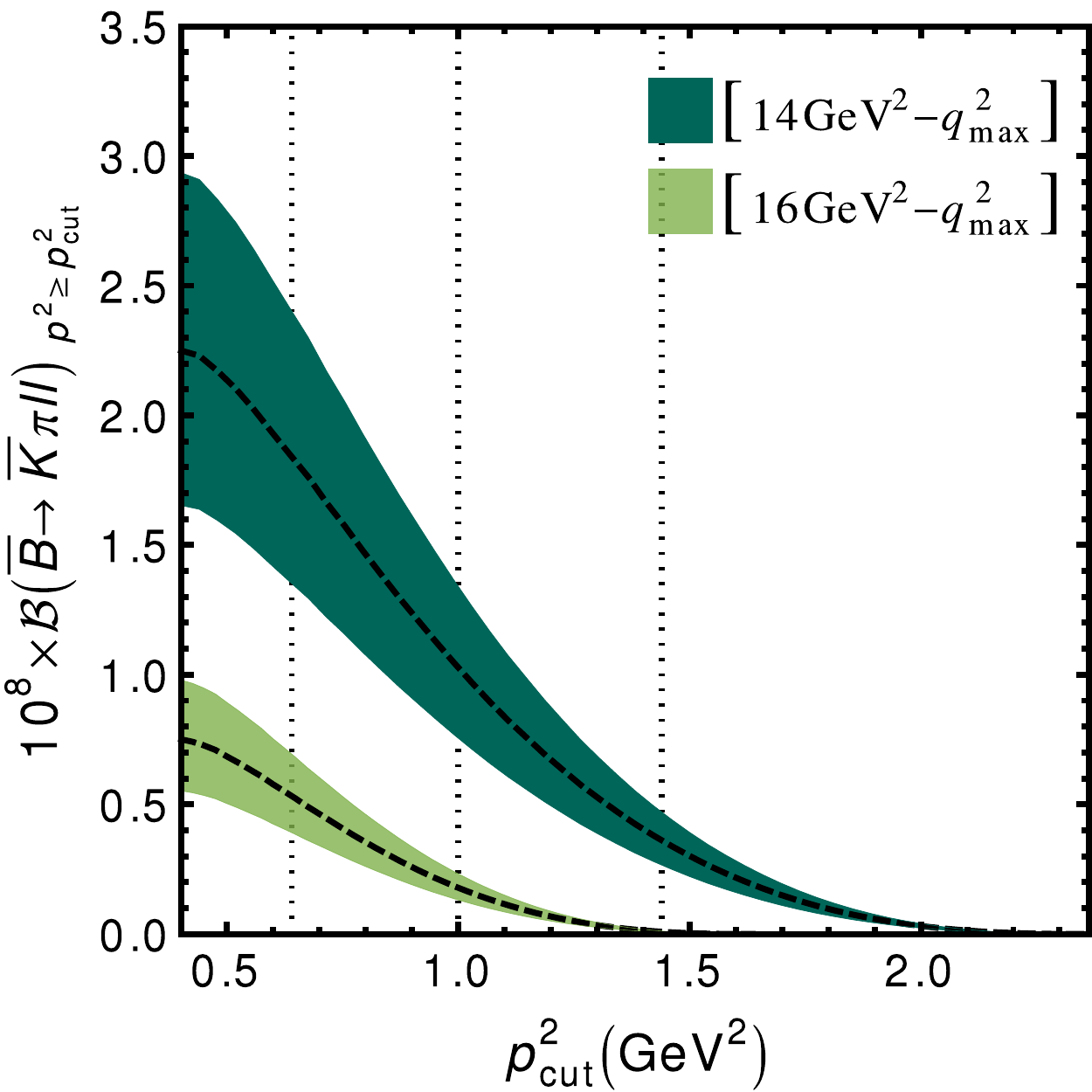}\qquad\quad
\includegraphics[height=0.4\textwidth]{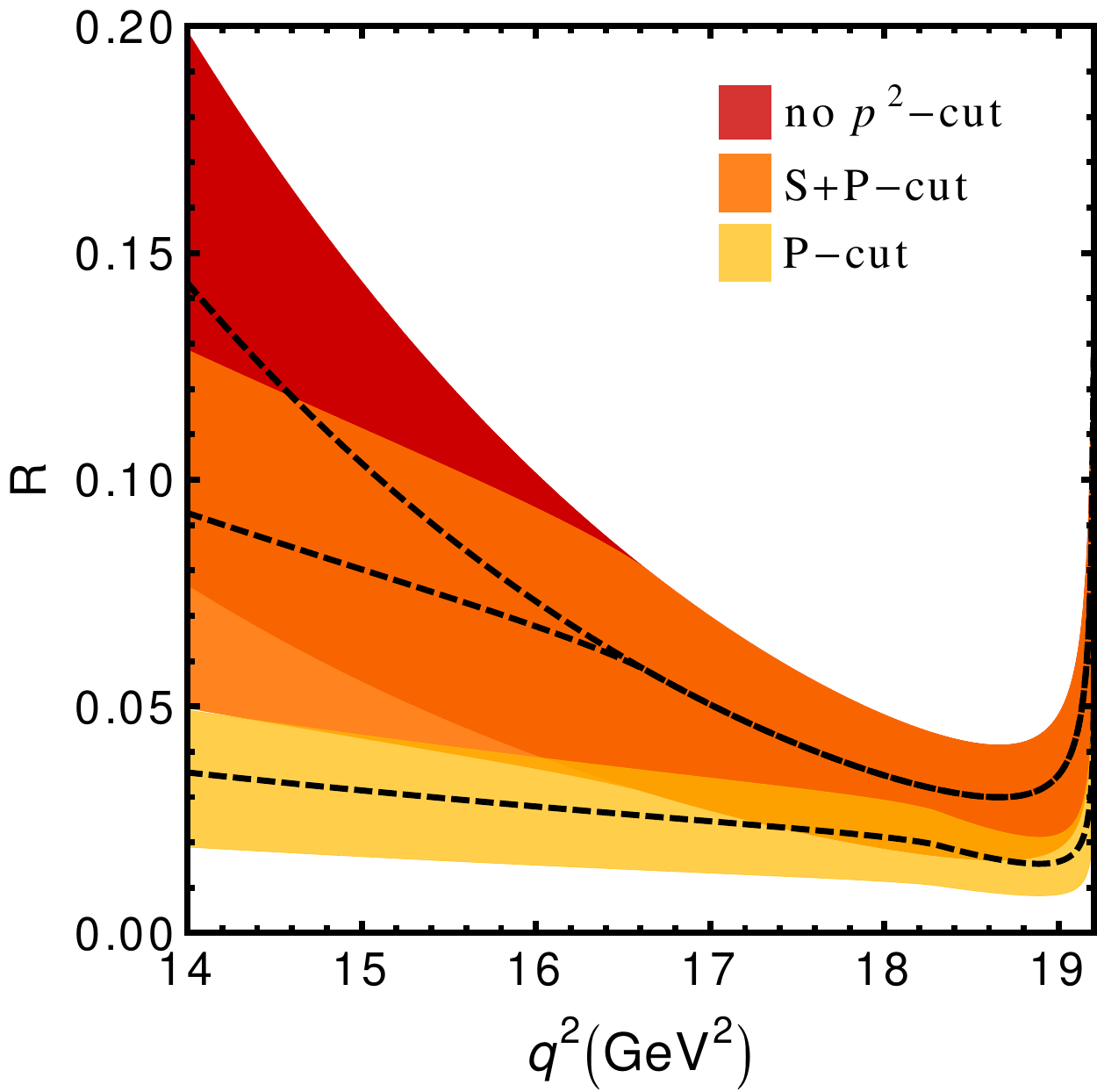}
}
\caption{\small Left: branching fraction of non-resonant $\bar B \to \bar K \pi \ell \ell$ decays in the SM, integrated over the low-recoil region from $(14,16) \, \mbox{GeV}^2-q^2_{\rm max}$ as a function of the lower $p^2$-integration boundary, as defined in Eq.~\eqref{eq:BRp2cut}. The vertical lines indicate (from left to right) the lower and upper bound for the P window, and the upper cut for the S+P window.
Right: $R=(d{\cal{B}}(\bar B \to \bar K  \pi \ell \ell)/dq^2)/(d{\cal {B}}(\bar B \to \bar K^* \ell \ell)/dq^2)$ in the SM basis for the three $p^2$-regions of interest.  
In the ratios the short-distance coupling $\rho_1(q^2)$ cancels, see Eq.~(\ref{eq:IopeSM}). The endpoint behavior is discussed in the text. 
In both plots dashed lines indicate central values of the input parameters.
 $\bar B \to \bar K \pi$ form factors are taken from HH$\chi$PT,  Eq.~(\ref{eq:ffinput}), and include parametric uncertainties only.
\label{fig:nr-cutBKpioverKstar}
}
\end{figure}

Outside the $K^*$ region, the integrated SM branching ratio in non-resonant decays is $10 \cdot  10^{-9}$ above and $4\cdot 10^{-9}$ below the signal window, and $4 \cdot 10^{-9}$ above the S+P window, see Fig.~\ref{fig:nr-cutBKpioverKstar} on the left, where the branching ratio for a variable cut in $p^2$ is shown for $q^2_{\rm min}=14,16\,{\rm GeV}^2$: 
\begin{align}
\label{eq:BRp2cut}
{\cal{B}}(\bar B \to \bar K \pi \ell \ell)_{p^2  \geq p^2_{\rm cut} }=  
\int_{q^2_{\rm min}}^{(m_B-m_K-m_\pi)^2}\!\!\! dq^2  \int_{p^2_{\rm cut}}   
\!\!\!dp^2\,\theta(\lambda)\, \frac{d^2{\cal{B}}}{d q^2 d p^2}(\bar B \to \bar K  
\pi \ell \ell)^{\rm SM} \, .
\end{align}

Uncertainties in the $\bar B \to \bar K \pi \ell \ell$ decay distributions stem from the following sources: 
{\it i)} the short-distance quantity $ \rho_1^{\rm SM}$, 
{\it ii)} parametric uncertainties, \emph{i.e.} $V_{ts}^* V_{tb}$ and the $B$-meson lifetime, 
{\it iii)} subleading $1/m_b$ corrections, and 
{\it iv)} form factors (parametric, especially $g$, and systematic, from higher order HH$\chi$PT corrections, which would induce strong phases, and $SU(3)$ breaking).
The latter ({\it iv}), which presently dominates the uncertainty budget, can be reduced in the future through improved determinations of the form factors $w_\pm,h$. We recall that we employ the leading order HH$\chi$PT results including some part of the $SU(3)$-corrections by
using  physical values for the decay constants, as detailed in Appendix \ref{app:input}. Remaining systematic uncertainties, which are not included in the numerical estimates in this work, can be at order few $10 \%$ in the form factors, making  further  determinations from other means desirable.
The uncertainties in $ \rho_1^{\rm SM}$, which is known at next-to-leading order in QCD, have been studied in detail in Ref.~\cite{Bobeth:2010wg} and amount to about 3~\%. 
The contributions to the uncertainty in the branching ratio from the CKM factors and the lifetime are about 5 \% and 0.5 \%, respectively. 
Further effects from charmonium resonances $\bar B \to  (\psi^{\prime ..} \to \ell \ell)  \bar K \pi$, not captured by insufficient bin size  or unfortunate bin boundaries,
are  known from related $\bar B \to \bar K^{(*)} \ell \ell$ studies and apply analogously. Factorizable resonance effects drop out in several ratios \cite{Hiller:2013cza}, see
\cite{Lyon:2014hpa} for a recent study.
The latter works found also sizable non-factorizable charmonium contributions in 
$ B \to K \mu \mu$ data. If and at which level this implies corrections to universality
has to be settled experimentally in the future, {\it e.g.}, with $\bar B \to \bar K^* \ell \ell$ angular
analysis.
The result for the fully integrated branching ratio of the non-resonant decay in the SM at low recoil reads
\begin{align} \label{eq:full-nonresoSM}
  10^8\cdot \int_{14 \, \mbox{\scriptsize  GeV}^2}^{q^2_{\rm max}} 
    \dd q^2 \frac{\dd {\cal {B}}^{\rm SM}}{\dd q^2} (\bar B \to \bar K \pi \ell \ell)& = 
   2.22\, {^{+0.66}_{-0.56}}\Big|_{\rm g}\, 
           {\pm 0.12}\Big|_{\rm CKM} \, {^{+0.07}_{-0.06}}\Big|_{\rm SD}.
\end{align}

The 'signal-to-background' ratio $R=(d{\cal{B}}(\bar B \to \bar K  \pi \ell \ell)/dq^2)/(d{\cal {B}}(\bar B \to \bar K^* \ell \ell)/dq^2)$ in the SM basis is shown in Fig.~\ref{fig:nr-cutBKpioverKstar} on the right. 
Note that here the short-distance coupling $\rho_1(q^2)$ drops out, as shown in Eq.~(\ref{eq:IopeSM}), and the results hold model-independently. 
The distributions for $\bar B \to \bar K^* \ell \ell$ can be obtained from the general formula, Eq.~\eqref{eq:full}, by projecting out the spin-1 component as shown in Appendix~\ref{app:resonant}.
These agree with previous findings on the angular distributions in \cite{Bobeth:2010wg}.
The form factors for $\bar B \to \bar K^*$  used in this work are taken from \cite{Ball:2004rg} as compiled in  \cite{Bobeth:2010wg}, and
employ an uncertainty estimate for the ratios $V/A_1$ of $8 \%$ and $A_2/A_1$ of $10 \%$ from \cite{Hambrock:2013zya}.
The ratio $R$
diverges at the $K^*$ endpoint in the zero-width approximation, $q^2=(m_{B}- m_{K^*})^2=19.21 \, \mbox{GeV}^2$, but is regularized in finite width by replacing the phase space factor $\lambda_{K^*}$ by $\lambda$.
The remaining theoretical uncertainties in $R$ stem from the heavy-quark expansion and form factors in the numerator and denominator, added in quadrature, where the latter could be reduced by an improved (and perhaps even combined) calculation of the $\bar B \to \bar K^*$ and
$\bar B \to \bar K \pi$ form factors. The $p^2$ cuts are seen to be rather efficient in suppressing the non-resonant decays  over the whole low-recoil $q^2$ region. However, a contribution of several percent remains, even for the $K^*(892)$ signal region. As will be seen in Section~\ref{sec:kstarreso}, this thereby constitutes the dominant background.

\begin{figure}[ht]
\centering{
\includegraphics[width=0.4\textwidth]{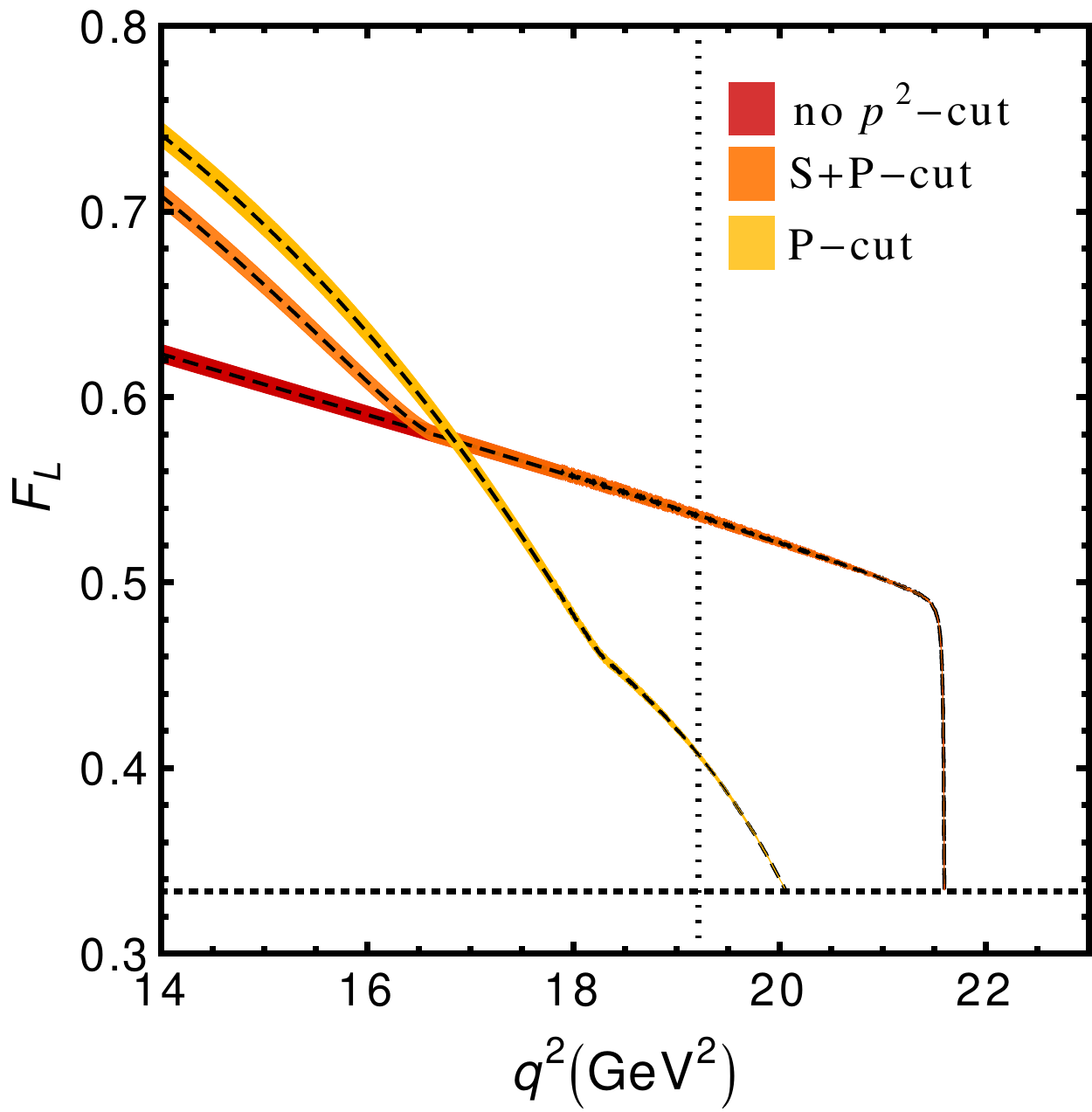}
}
\caption{\small The angular observable $F_L$ for $\bar B \to \bar K \pi \ell \ell$ with the three different $p^2$ cuts
in the SM basis.
Dashed lines are for central values of the input parameters. The short-distance coupling $\rho_1(q^2)$ drops out in the SM basis, see Eq.~(\ref{eq:IopeSM}). The horizontal dotted line marks the endpoint prediction $F_L=1/3$, see Section  \ref{sec:ep}, the vertical one the endpoint of the $K^*(892)$ distribution. 
 $\bar B \to \bar K \pi$ form factors are taken from HH$\chi$PT,  Eq.~(\ref{eq:ffinput}), and include parametric uncertainties only.
\label{fig:BKpi-obs}
}
\end{figure}

The impact of the different $p^2$ cuts on angular observables is shown exemplarily in Fig.~\ref{fig:BKpi-obs} for $F_L$, the fraction of longitudinal transversity states.
In the SM basis at low recoil short-distance couplings cancel and $F_L$ (locally) measures form factor ratios. 
In $F_L$ also form factor uncertainties cancel; 
specifically, for HH$\chi$PT form factors  this concerns the decay constants and to a  large extent the coupling $g$, although it enters the form factors with different powers, cf. Eq.~(\ref{eq:ffinput}), and this cancellation is not perfect.
Note also that systematic uncertainties in the form factors have not been included. We expect, however, that some cancellations take place in ratios.
Nevertheless, it shows that appropriately constructed observables can be predicted with much higher precision than the differential rate.
The endpoint value of non-resonant decays equals $1/3$ as predicted, yet the very same
$K^*(892)$ endpoint prediction is contaminated from non-resonant backgrounds because
$F_L=1/3$ is assumed at different values of $q^2$. The steep approach towards the maximal non-resonant $q^2$ is caused by $\lambda_p$, which vanishes at this point.

\subsection{Angular Analysis\label{sec:angularpheno}}

Approximating the non-resonant distributions by their S, P, D partial waves is a useful approximation, especially when discussing interference with resonant contributions. 
The following aspects of the resulting distribution, as given in Eq.~(\ref{eq:full}), are addressed:
Impact of non-resonant contributions on the $\bar B \to \bar K^* \ell \ell$ angular coefficients  (Subsection \ref{sec:BGD}), measurement of strong phases, specifically  impact of relative phases from the $K^*$ overlapping with 
the non-resonant contributions 
(Subsection \ref{sec:strong}) and predictions for contributions to $\bar B \to \bar K^* \ell \ell$ SM null tests (Subsection \ref{sec:null}).
We treat the $K^*(892)$ in zero-width approximation.
A full study of the physics reach of the angular analysis, including correlations, beyond zero-width-$K^*$ or global fits, is beyond the scope of the present work.

\subsubsection{Background to the angular coefficients in $\bar B\to \bar K^*\ell\ell$ \label{sec:BGD}}

\begin{figure}[ht]
\centering{
\includegraphics[height=0.4\textwidth]{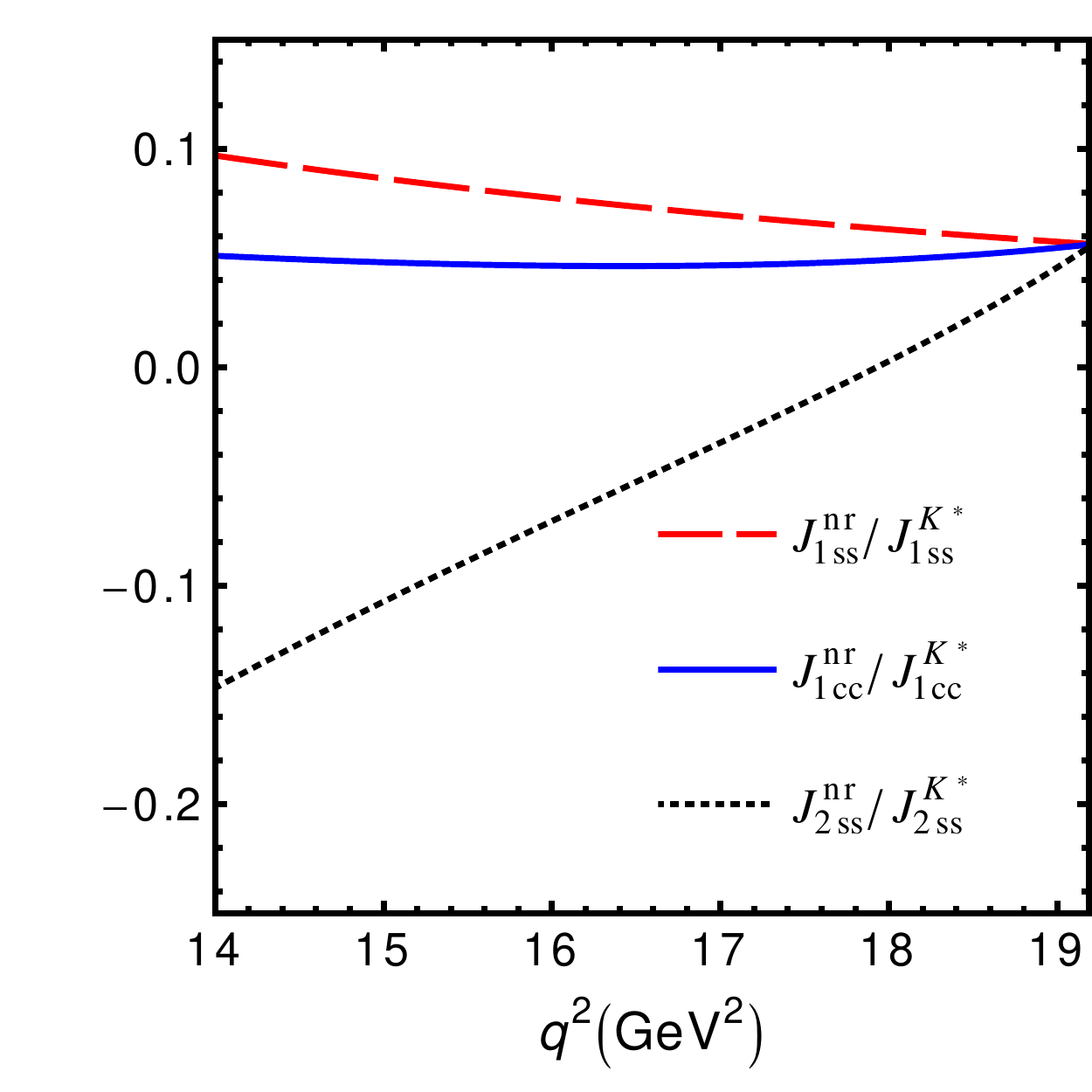}\qquad\quad
\includegraphics[height=0.4\textwidth]{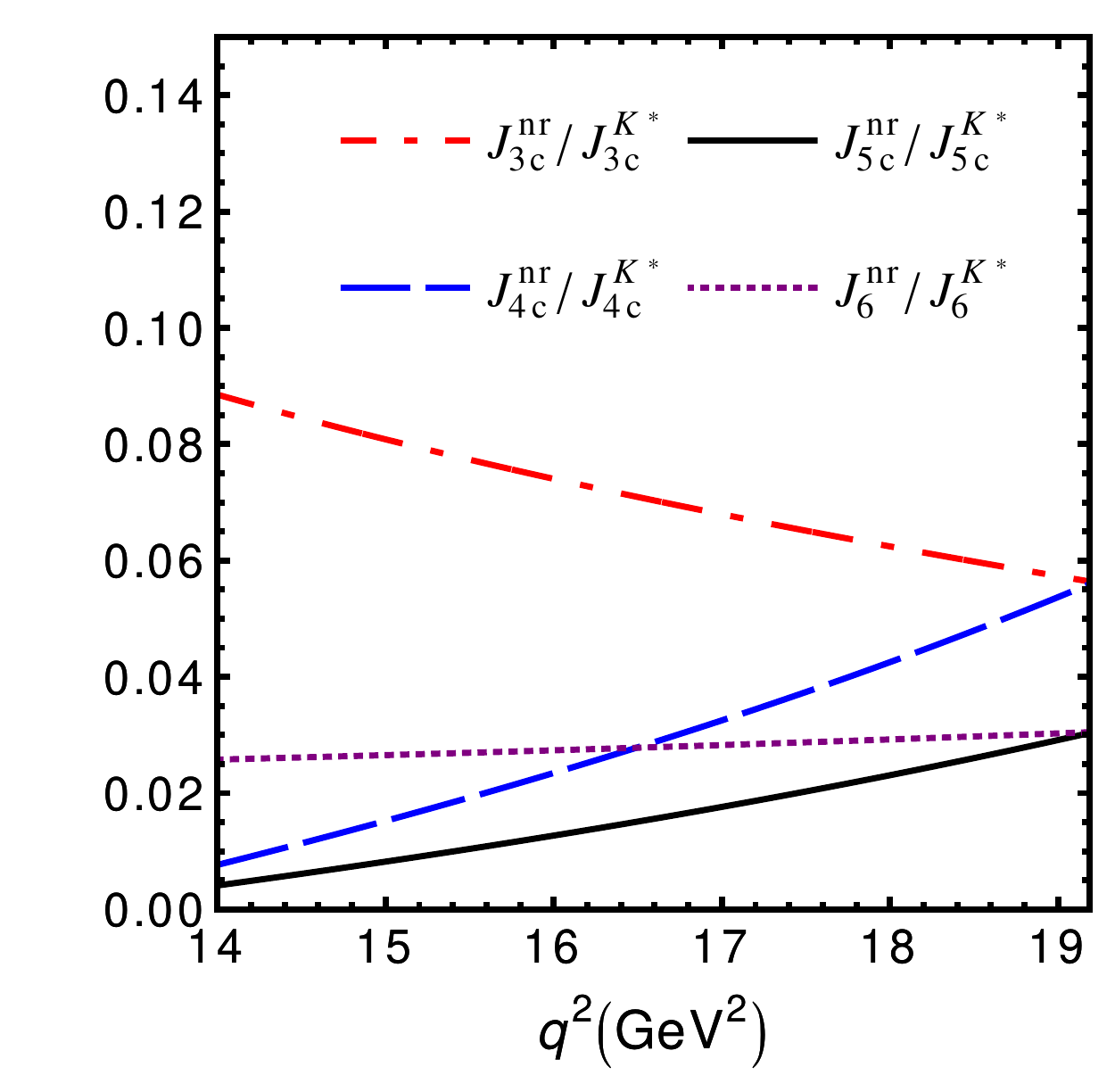}
}
\caption{\small The ratio $J_{ix}^{\rm nr}/J_{ix}^{K^*}$ for $p^2=m_{K^*}^2$ in the SM basis for 
$J_{1cc}(=J_{2cc})$, $J_{1ss}$ and $J_{2ss}$ (left plot) and
$J_3$, $J_{4c}$, $J_{5c}$ and $J_6$ (right plot). Form factor uncertainties are not included. 
\label{fig:ratioJ}
}
\end{figure}

In order to estimate the influence of the non-resonant contributions more generally, the relative contributions in the angular coefficients of $\bar B\to \bar K^*\ell\ell$, $J_{ix}^{\rm nr}/J_{ix}^{K^*}$, are calculated underneath the mass peak of the $K^*$ at low recoil in the SM basis, from which the influence on all observables in the $\bar B\to \bar K^*\ell\ell$ analyses can be estimated.
In this basis the short-distance physics cancels in these ratios.
The results are shown in Fig.~\ref{fig:ratioJ}; the corresponding curves for $J_{7c,8c,9}$ are discussed in the next subsection, as for those $J^{K^*}$ are null tests in the lowest order OPE.
All ratios are at the few
percent level, except for the ones of $J_3$ and $J_{iss}, i=1,2$, which can be larger
 and increase towards lower values of $q^2$. 
For  $J_3$ this happens because $J_3^{K^*}$ vanishes  in the limit $f_\parallel(K^*) \sim f_\perp(K^*)$. The latter can be understood in terms of helicity conservation \cite{Burdman:2000ku} and, while being 
formally a feature at {\it large} recoil, starts setting in already
at the lower $q^2$-end of the low-recoil region. The effect in $J_{iss}, i=1,2$ stems from the numerator and is predominantly caused by the sizable non-resonant S-wave component in $F_{0}$, see Fig.~\ref{fig:aSPD}.

These ratios are typically of the same size as the relative contributions of non-resonant decays to the differential branching ratio, as shown in Fig.~\ref{fig:nr-cutBKpioverKstar}. However, some of the ratios can also be significantly larger, implying an even larger influence on other observables, rendering their inclusion mandatory.

\subsubsection{Probing strong phases  \label{sec:strong}}

\begin{figure}[ht]
\centering{
\includegraphics[width=0.4\textwidth]{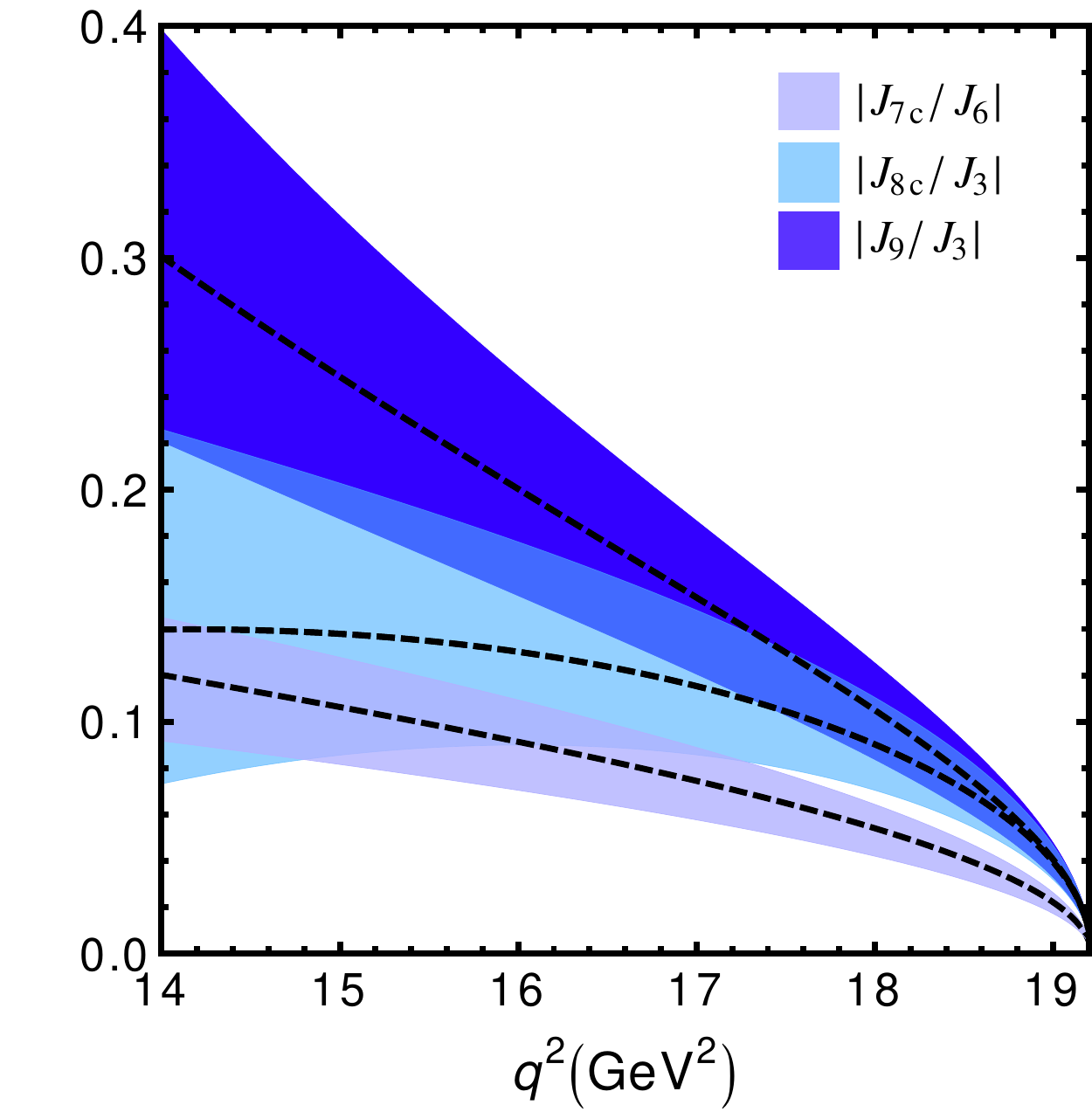}
}
\caption{\small Tests for relative strong phases: $|J_{7c}/J_6|$,  $|J_{8c}/J_3|$ and $|J_{9}/J_3|$ for $p^2=m_{K^*}^2$ and maximal relative strong phase $\delta_{K^*}=\pi/2$ in the SM basis, see text for details.
\label{fig:phases}
}
\end{figure}

The angular coefficients associated with $I_{7-9}$ are particularly sensitive to relative strong phases since they  vanish without the latter in the SM basis, as shown in Eq.~(\ref{eq:IopeSM}). The coefficients associated with
$I_7$ vanish even in the more general SM+SM$^\prime$ basis in this case, 
as can be seen from Eq.~(\ref{eq:Iope}). This offers opportunities to probe the relative strong phases between
resonant and non-resonant contributions, and between different resonances.
For interference effects to be sizable, the resonant P-wave contribution from $\bar B\to \bar K^*\ell\ell$ has to be involved. 
The following observables therefore probe the interference of the non-resonant P-wave with the $K^*$ contribution cleanly. In particular, the normalization to $J_{3,6}$ guarantees that is no additional S- or D-wave ``pollution'' in the denominator.
\begin{align}
\frac{J_{7c}}{J_6} & \simeq \frac{{\rm Im}(F_{0P} F_{\parallel P}^*)}{{\rm Re}(F_{\parallel P} F_{\perp P}^*)} \, , \\
\frac{J_{8c}}{J_3} & \simeq 2 \frac{{\rm Im}(F_{0P} F_{\perp P}^*)}{|F_{\perp P}|^2-| F_{\parallel P}|^2} \, ,\\
\frac{J_{9}}{J_3}  & = 2 \frac{{\rm Im}(F_{\perp P} F_{\parallel P}^*)}{|F_{\perp P}|^2-| F_{\parallel P}|^2}  \, .
\end{align}
In Fig.~\ref{fig:phases} the absolute values of these ratios are shown in magnitude for a maximal strong-phase difference, in order to see how large they can become. 
The curves would vanish for $\delta_{K^*}=0,\pi$. We recall that in general the phase is expected to vary over the phase space.
It is seen that contributions can be sizable and can be benefited from in an angular analysis. Note that all these ratios depend dominantly on \emph{one} relative strong phase between the $K^*$ and the non-resonant contribution, $\delta_{K^*}$. Its extraction is possible even in the presence of an extended operator basis, where the new combinations of Wilson coefficients, $\delta \rho$ and $\rho_2^-$, appear which are discussed in Section~\ref{sec:sd}.
If $\delta \rho$ and $\rho_2^-$ are complex, potentially further contributions to $J_{8x}$ and $J_{9x}$ arise, while $J_{7x}$ still requires a finite relative strong phases to be non-zero.
In that case it useful that different combinations of form factors and these coefficients enter the angular coefficients $J_{(7-9)x}$, see Appendix~\ref{app:resonant}. 
Note again that $I_9$ is free from S-wave contributions.

\subsubsection{$\bar B \to \bar K^* \ell \ell$ SM null tests \label{sec:null}}

\begin{figure}[ht]
\centering{
\includegraphics[width=0.4\textwidth]{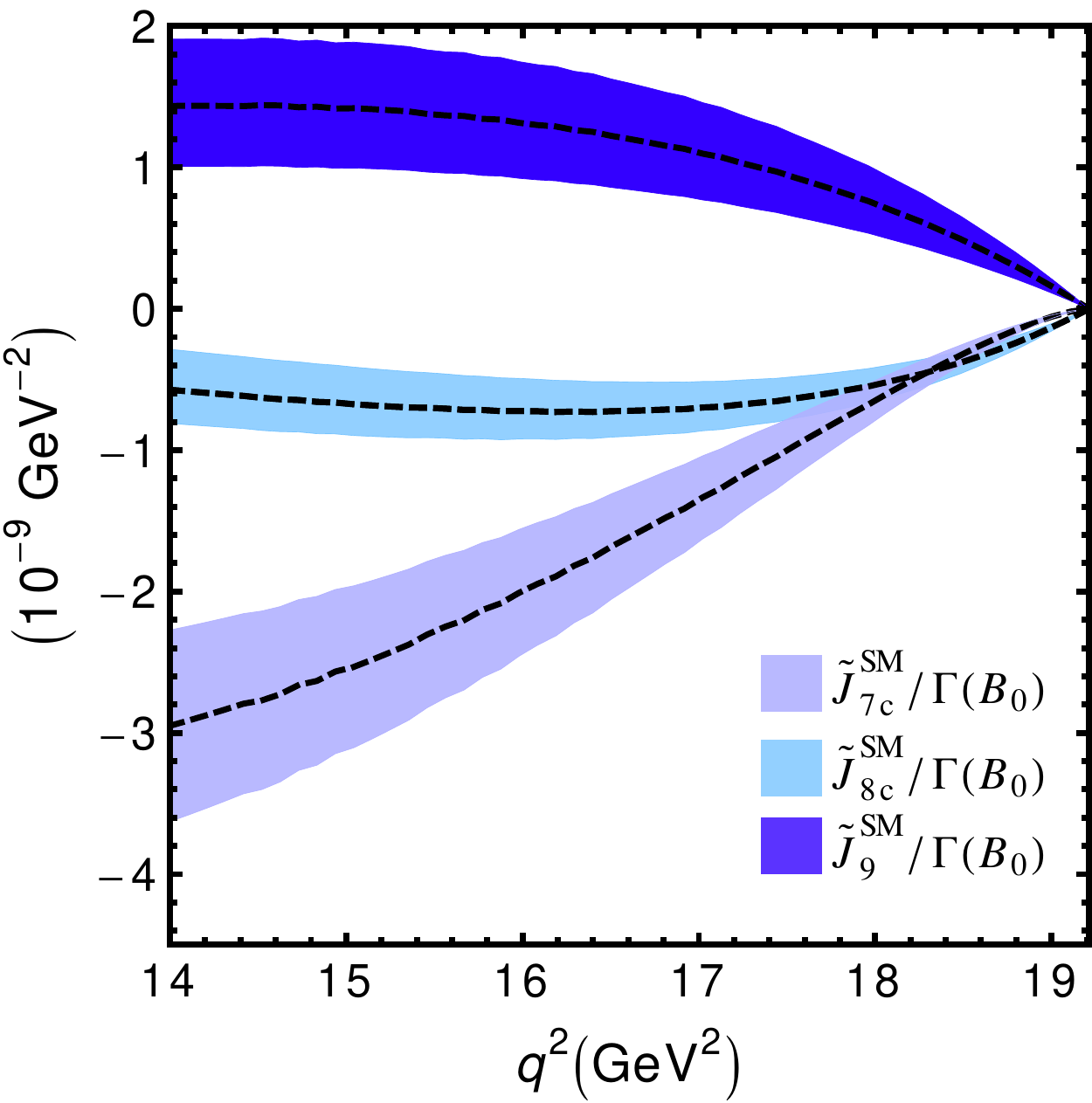}
}
\caption{\small SM predictions for $\tilde J_{7c,8c,9}^{\rm SM}/\Gamma(B_0)$
for  maximal relative strong phase $\delta_{K^*}=\pi/2$.
The observables would vanish for a pure resonant contribution or $\delta_{K^*}=0,\pi$.
\label{fig:789SM}
}
\end{figure}

The coefficients discussed in the last paragraph vanish in the SM basis for a purely resonant decay $\bar B\to \bar K^*\ell\ell$, thereby providing null tests of the standard analysis of this mode.
In Fig.~\ref{fig:789SM}, they are shown as a function of the  
dilepton invariant mass squared, for $p^2=m_{K^*}^2$ in the zero-width approximation and normalized to the total width $\Gamma(B_0)$:
\begin{align}
\tilde J_{7c}^{\rm SM} & =-\rho_2^{\rm  
SM}   \,{\rm Im}\left(F_{0P} F_{\parallel P}^*\right) \, ,\\
\tilde J_{8c}^{\rm SM} & = 
\frac{\rho_1^{\rm SM}}{4}   \,{\rm Im}\left(F_{0P} F_{\perp P}^*\right) \, ,\\
\tilde J_{9}^{\rm SM} & = 
\frac{\rho_1^{\rm SM}}{4}   \,{\rm Im}\left(F_{\perp P} F_{\parallel P}^*\right)  
\, .
\end{align}
The effect in $J_{7c}$ is the largest among the observables studied. When compared to the
differential branching ratio of $\bar B \to \bar K^* \ell \ell$  in the SM, the induced change $\tilde J_{7c}^{\rm SM}/\Gamma(B_0)$ can be up to $\sim 5 \%$ in magnitude.

\subsection{Resonant S-wave contributions to $\boldsymbol{\bar B \to \bar K\pi \ell \ell}$  \label{sec:kstarreso}}

Semi-leptonic decays to the $\bar K\pi\ell\ell$ final state contain contributions from decays proceeding via kaon resonances. The relevant states are detailed in 
Table~\ref{tab:kpi}.
\begin{table}[ht]
\begin{center}
\begin{tabular}{c|c|c|c|c}
\hline\hline
    &     $J^P$ &mass [MeV] & width  [MeV]  & branching ratio to $\bar K \pi$ \\\hline\hline
$\kappa(800) $ & $0^+$& $ 658 $ & $557 $& $\sim 100$ \% \\
$K^{*0}(892) $ & $1^-$ & $ 895.8 $ & $47.4 $& $\sim 100$ \%\\
$K^{*}(1410) $ & $1^-$ & $ 1414 $ & $232 $& $\sim 7$ \%\\
$K^*_0(1430)$ & $0^+$& $ 1425 $ & $270 $ & $\sim 100$ \%\\
$K^{*0}_2(1430)$ & $2^+$& $ 1432 $ & $109 $& $\sim 50$ \%\\
$K^{*}(1680) $ &$1^-$ & $ 1717 $ & $322 $& $\sim 39$ \%\\
$K^{*}_3(1780) $ & $3^-$ & $ 1776 $ & $159 $& $\sim 19$ \%\\
  \hline\hline
\end{tabular}
\end{center}
\caption{\label{tab:kpi}Selected states decaying to $\bar K \pi$  \cite{PDG}. Data on $\kappa(800)$ from \cite{DescotesGenon:2006uk}.
}
\end{table}
The angular distributions for semi-leptonic $\bar B$ to spin-0 kaons can be obtained from
the  general formula, Eq.~\eqref{eq:full}, by projecting out the spin-0 component, as shown in Appendix \ref{app:resonant}.

The $p^2$ line shape of the resonant S-wave contributions to \BKpill{} can be described phenomenologically by the coupled Breit-Wigner formalism set out in Ref.~\cite{Becirevic:2012dp},
\begin{align} \label{eq:bt}
BW_S(p^2)&  ={\cal{N}}_{S} \left[  \frac{-g_\kappa}{(m_\kappa-i \Gamma_\kappa/2)^2-p^2}+ \frac{1}{(m_{K^*_0}-i \Gamma_{K^*_0}/2)^2-p^2}\right] \, ,
\end{align}
with the normalization factor ${\cal{N}}_{S}$ fixed by
\begin{align}
 \int_{-\infty}^{\infty} dp^2 | BW_S(p^2)|^2= 1 \, .
\end{align}
We stress that we are not aiming at a first-principle description of the $\kappa$ line shape but rather employ (\ref{eq:bt}) as a simple, data-based parameterization.

The parameter $g_\kappa$ is complex in general and the data on $(\bar K\pi)$ line shapes is well approximated at least in the  'signal' window by the parameter values
$|g_\kappa| \lesssim 0.2$, $\pi/2 \lesssim \arg g_\kappa \lesssim \pi$.\footnote{We thank Damir Becirevic for communication on this point.}
Since the resonances are considered within this formalism at finite width, the appropriate phase-space factor is the function $\lambda$ introduced in Section~\ref{sec:matrix}.
A comparison of different line shapes in a wider
$p^2$ region can be seen in Ref.~\cite{Doring:2013wka}.\footnote{The dotted and dashed curves in Fig.~7 of Ref.~\cite{Doring:2013wka}  are labeled erroneously and should be interchanged. We thank the authors  for confirmation.} After taking into account that the 
agreement of the parametrization in Eq.~\eqref{eq:bt} with the other ones discussed there improves for the value $\arg g_\kappa=\pi/2$ (this work, see below)
when compared to $\arg g_\kappa \simeq 0$, used in Ref.~\cite{Doring:2013wka}, the difference between the predictions is at  the 30\% level. This can be taken as an indication for the size of uncertainties in the 
resonant $\bar K \pi$ S-wave background.

\begin{figure}[ht]
\centering{
\includegraphics[width=0.4\textwidth]{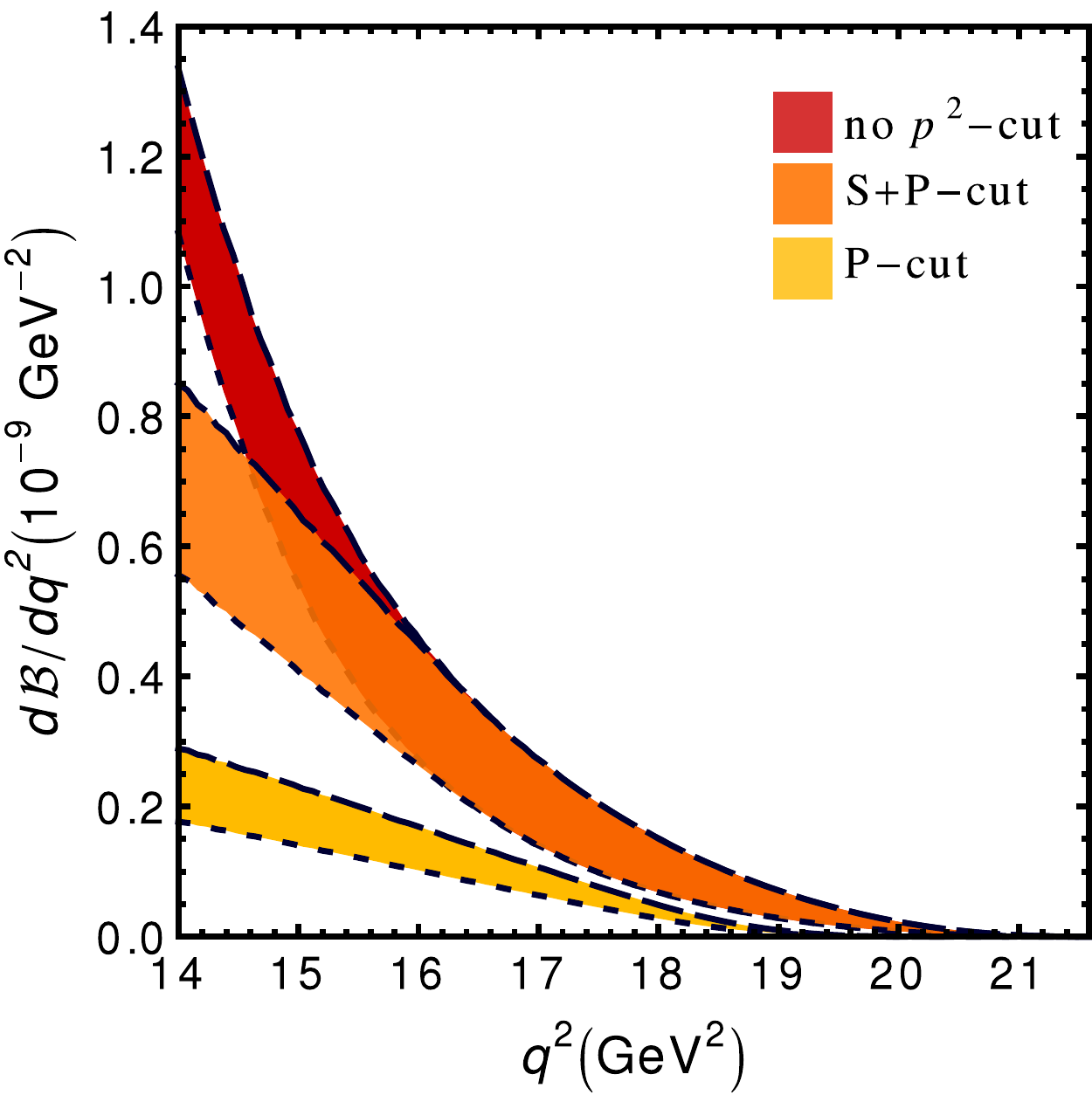}\qquad\quad 
\includegraphics[width=0.41\textwidth]{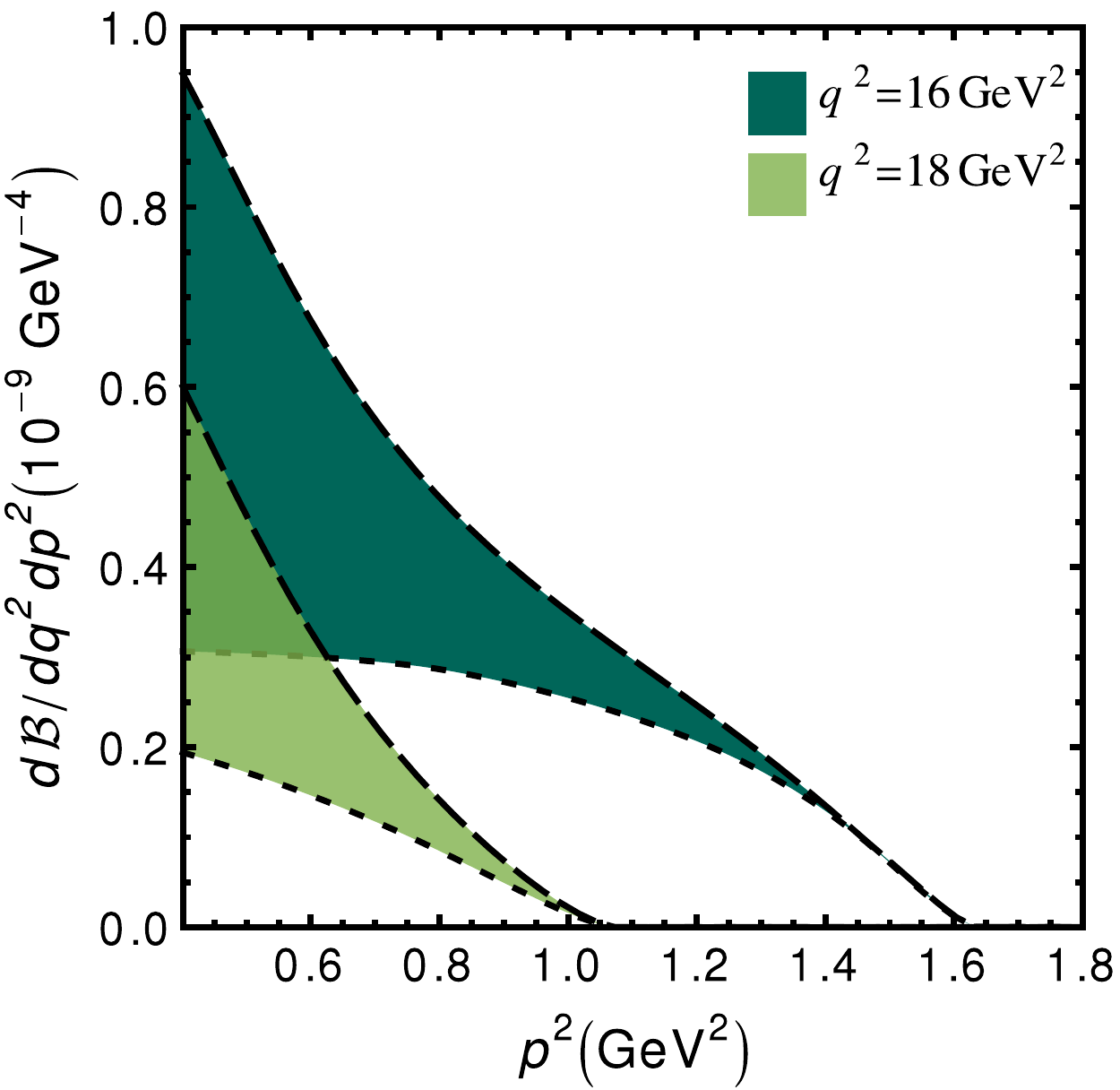} 
}
\caption{\small Differential branching fraction of resonant S-wave 
$\bar B \to ( \bar K_0^*(1430) +\kappa(800) )\ell \ell$ decays using the line shape given in Eq.~(\ref{eq:bt}) \cite{Becirevic:2012dp},
as a function of $q^2$ for the three $p^2$ regions (left) and for $p^2$ at two given values of $q^2$ (right).
The branching fractions are given using the numerical input in Table~\ref{tab:input},  
QCDSR form factors \cite{Aliev:2007rq}, $|g_\kappa|=0,0.2$ (dotted and dashed lines, respectively) and $\,\arg g_\kappa=\pi/2$.  
Form factor uncertainties are not included.
\label{fig:BT}
}
\end{figure}  

\begin{figure}
\centering{
\includegraphics[height=0.4 \textwidth]{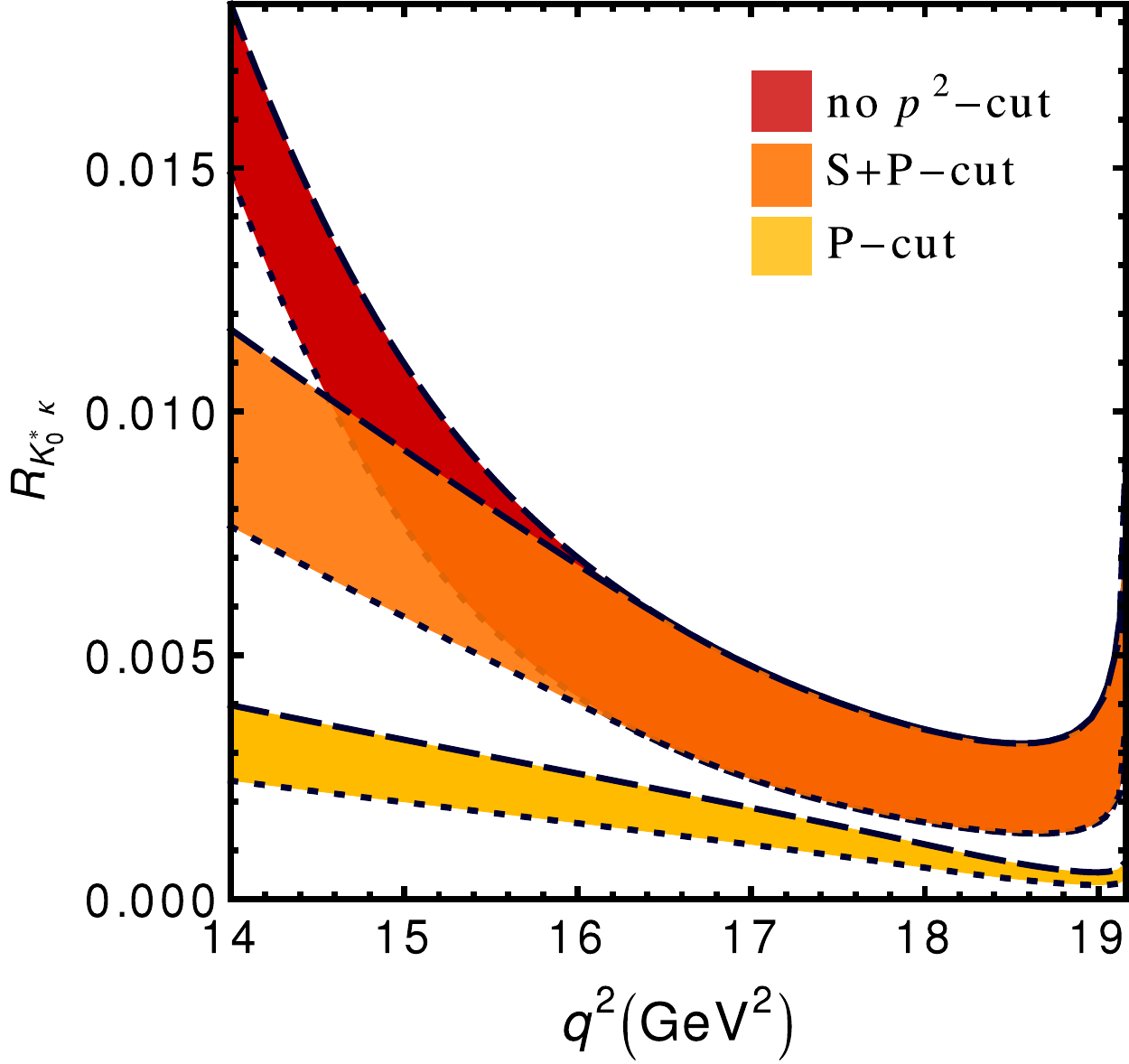}
}
\caption{\small $R_{K_0^*\kappa}=(d{\cal{B}}(\bar B \to (\bar K_0^*(1430)+\kappa(800)) \ell \ell)/dq^2)/(d{\cal {B}}(\bar B \to \bar K^* \ell \ell)/dq^2)$ for the resonant S-wave contributions, using the line shape given in Eq.~(\ref{eq:bt}) for $\arg g_\kappa=\pi/2$, $|g_\kappa|=0,0.2$ (dotted and dashed lines, respectively) (form factor uncertainties are not included) in the SM basis for the three $p^2$-cuts.  
In the ratios the short-distance coupling $\rho_1(q^2)$ cancels, see Eq.~(\ref{eq:IopeSM}). The endpoint behavior is discussed in the text. \label{fig:resSwaveoKstar}}
\end{figure}

The differential branching fraction of \BKpill, where the $\bar K\pi$ comes from a $K_0^*(1430)$ or $\kappa(800)$ state in the low-recoil region, is shown in Fig.~\ref{fig:BT}.
The value of the parameters chosen are $|g_{\kappa}| = 0.2$ and $\arg(g_\kappa)=\pi/2$, in order to maximize the S-wave distribution in the region preferred by experimental data.
The dotted curves correspond to $g_\kappa=0$, {\it i.e.} no resonant contribution from the $\kappa(800)$. When compared to the non-resonant differential branching fractions shown in Fig.~\ref{fig:BKpi}, the magnitude of the resonant differential branching fractions is 
subdominant to the non-resonant one. This is further illustrated in Fig.~\ref{fig:resSwaveoKstar}, which is the analogue to Fig.~\ref{fig:nr-cutBKpioverKstar} for the resonant S-wave contributions. Again form-factor uncertainties from the numerator and denominator enter and are added in quadrature.
Form factors for $B \to K^*_0(1430)$ are taken from QCD sum rules (QCDSR)~\cite{Aliev:2007rq}. A calculation within the light front quark model \cite{Chen:2007na} yields consistent values at low recoil. The form factor estimates are considered rather uncertain already at large momentum transfer; additionally, they have to be extrapolated to the large recoil region. Note that they depend on the interpretation of the $\kappa(800)$ resonance; Ref.~\cite{Aliev:2007rq} assumes that the $K_0^*(1430)$ is the lowest scalar $s\bar q$ resonance. A comparable estimate in pQCD~\cite{Li:2008tk} yields very large values for the form factor at low recoil, which we consider unrealistic.
It can be seen that resonances with higher mass and $J \neq 1$, including the $K_2^*(1430)$, have less overlap with the $K^*(892)$ and/or a smaller rate to $\bar K \pi$, as shown in Table \ref{tab:kpi}, and consequently their impact on angular analyses is even smaller. 

The $J=1$ resonances with higher mass and their resulting P-wave
contributions do not change the structure of the $\bar B \to \bar  K ^*\ \ell \ell$ angular distribution,
while contributing to the generalized transversity form factors ${\cal{F}}_i \propto \sum F_{iP}$.
This way, the short-distance/long-distance separation with universal short-distance coefficients
of the low recoil region remains intact. Since ratios of form factors extracted from data \cite{Hambrock:2012dg}
are accessible only as superposition of $1^-$ states with relative strong phases, 
information on the $p^2$-dependence is necessary
before they can be compared to predictions for $\bar B\to\bar K^*$ from lattice QCD or sum rule calculations.

We conclude that the non-resonant decays form the largest part of the  
background in the $\bar B\to\bar K^*\ell\ell$ 'signal' window. While  
such effects are at the order of a few percent in the decay rate, their  
size  generically differs depending on the observable in question.  
Some angular observables have been studied in this regard in the  
previous subsection. In the future
the accuracy to which these effects can be predicted can be further  
improved with better knowledge of the $\bar B\to \bar K\pi$  form  
factors.
A sufficiently precise non-resonant  distribution may allow to  
circumvent sideband subtractions.
We stress  that this depends on the angular coefficient involved.
Note also that relative strong phases
signal interference from non-resonant or resonant sources and can  
quantify any such admixture.

\subsection{Non-resonant $\boldsymbol{\bar B_s \to \bar KK\ell \ell}$ decays \label{sec:phi}}
The decays \BsKKll{} are similar to the \BKpill{} ones and it is possible to obtain similar predictions for the corresponding non-resonant contributions. A main difference between the two decays is the narrow width of the resonant P-wave state, which allows for much tighter cuts to isolate the signal.
In order to understand the contributions to the experimental distributions, the following regions of $p^2$ are chosen:
\begin{itemize}
\item[--] Full phase space of the non-resonant decay:  $p^2_{\rm min} \equiv (2  
m_K)^2 \leq p^2 < (m_{B_s}- \sqrt{q^2})^2$, with endpoint
$q^2=q^2_{\rm max} \equiv (m_{B_s}-\sqrt{p^2_{\rm min}})^2=19. 18\,  
\mbox{GeV}^2$.
\item[--] P-wave 'signal' window: $1.01  \, \mbox{GeV}^2 \leq p^2 <  
1.06 \, \mbox{GeV}^2$,  corresponding to $m_\phi \pm 12 $ MeV  
\cite{Aaij:2013aln} and the endpoint $q^2=19.03 \, \mbox{GeV}^2$.
\item[--] S+P-wave 'total' window: $p^2_{\rm min} \leq p^2 < (m_\phi+  
50 \, \mbox{MeV})^2=1.14 \, \mbox{GeV}^2$,  and the endpoint $q^2_{\rm  
max}$.
\end{itemize}
Note that the endpoint of the signal decay \BsKKll{} is at $q^2=(m_{B_s}- m_\phi)^2=18.90 \, \mbox{GeV}^2$.

Another important difference between the decays \BsKKll{} and \BKpill{} is that there are
no low lying scalar $(\bar s s)$ mesons, as can be seen in Table~\ref{tab:ssbar}, which contribute to the signal window.
This is because
the low-mass $(\bar s s)$ mesons have either small branching ratios to $\bar K K$
or do not overlap significantly with the $\phi$.
In this regard, the  $\bar B_s  \to \phi  \ell \ell$ decay is cleaner than the $\bar  
B \to \bar K^*  \ell \ell$ one as the latter contains resonant  
backgrounds at low recoil from states such as the $\kappa(800)$.
Furthermore, there are  opportunities in $B_s$ decays due to the finite  
lifetime difference, including untagged CP-asymmetries related to  
$I_{5,6,8,9}$ \cite{Bobeth:2008ij}.

\begin{table}[ht]
\begin{center}
\begin{tabular}{c|c|c|c|c}
\hline\hline
    &   $J^P$ &  mass   [MeV] & width  [MeV] & branching ratio to $ 
\bar K K$ \\\hline\hline
$f_0(980)$	& $0^+$	& $990$				& $\sim 70$		& subdominant\\
$\phi $ & $1^-$ &$ 1019$	& 4 & 48.9 \% ($K^+ K^-$)\\
   $f_2(1270)$ &$2^+$ & 1275 & 185 & 4.6 \%\\
   $f_0(1370)$	& $0^+$	& $\sim 1350$	& $\sim 350$	& subdominant\\
   $f_2(1430)$	& $2^+$	& $\sim 1430$	& unknown		& needs confirmation\\
  $f_0(1500)$  &$0^+$ & 1505 & 109 &  8.6 \%\\
   $f_2^\prime(1525)$  &$2^+$ & 1525 & 75 &   89 \%\\
   \hline\hline
\end{tabular}
\end{center}
\caption{\label{tab:ssbar}
Available information for selected $(\bar s s)$ mesons decaying to $\bar K K$ \cite{PDG}.
}
\end{table}

   \begin{figure}[ht]
\centering{
\includegraphics[height=0.4\textwidth]{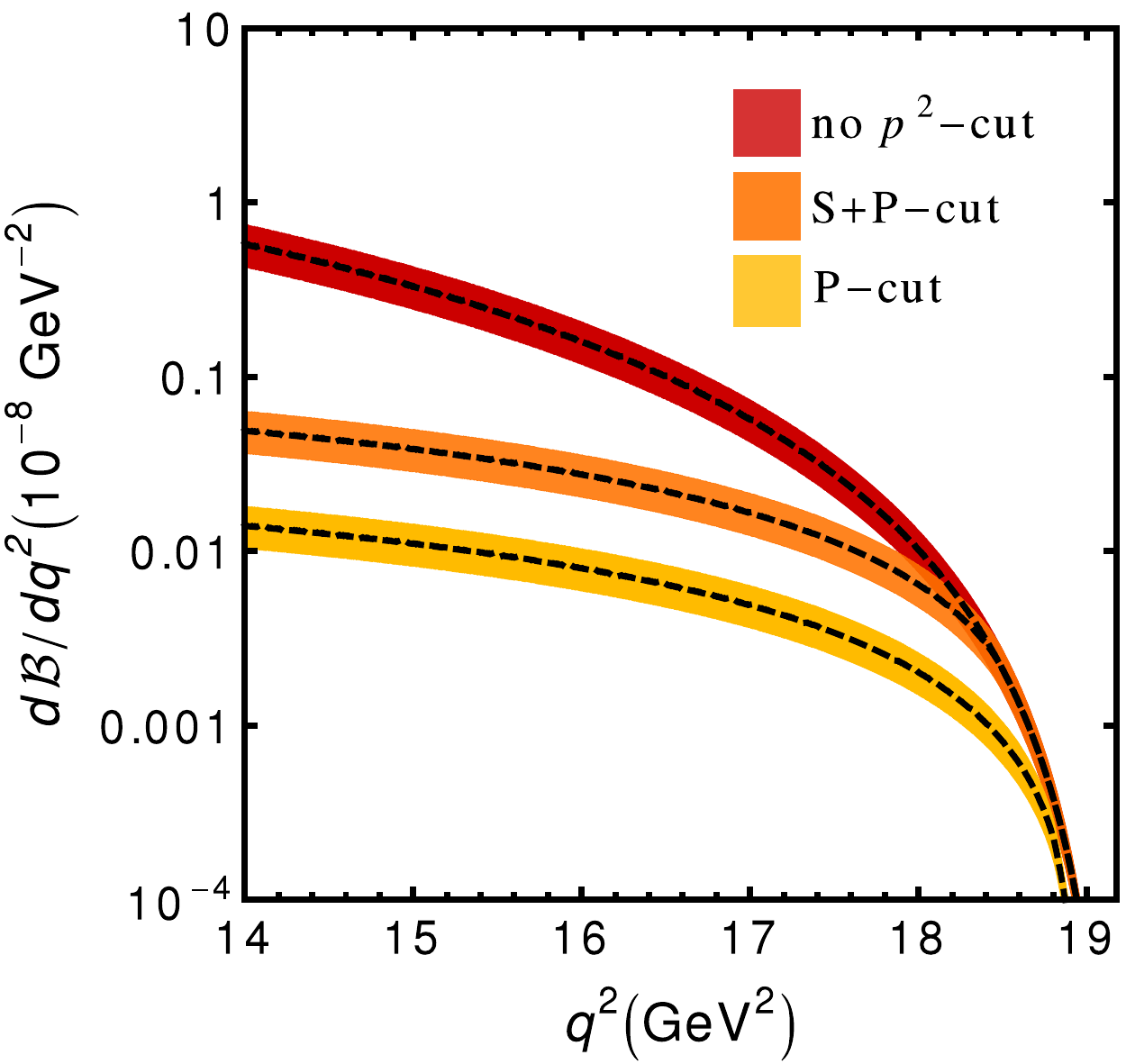}\qquad\quad
\includegraphics[height=0.4\textwidth]{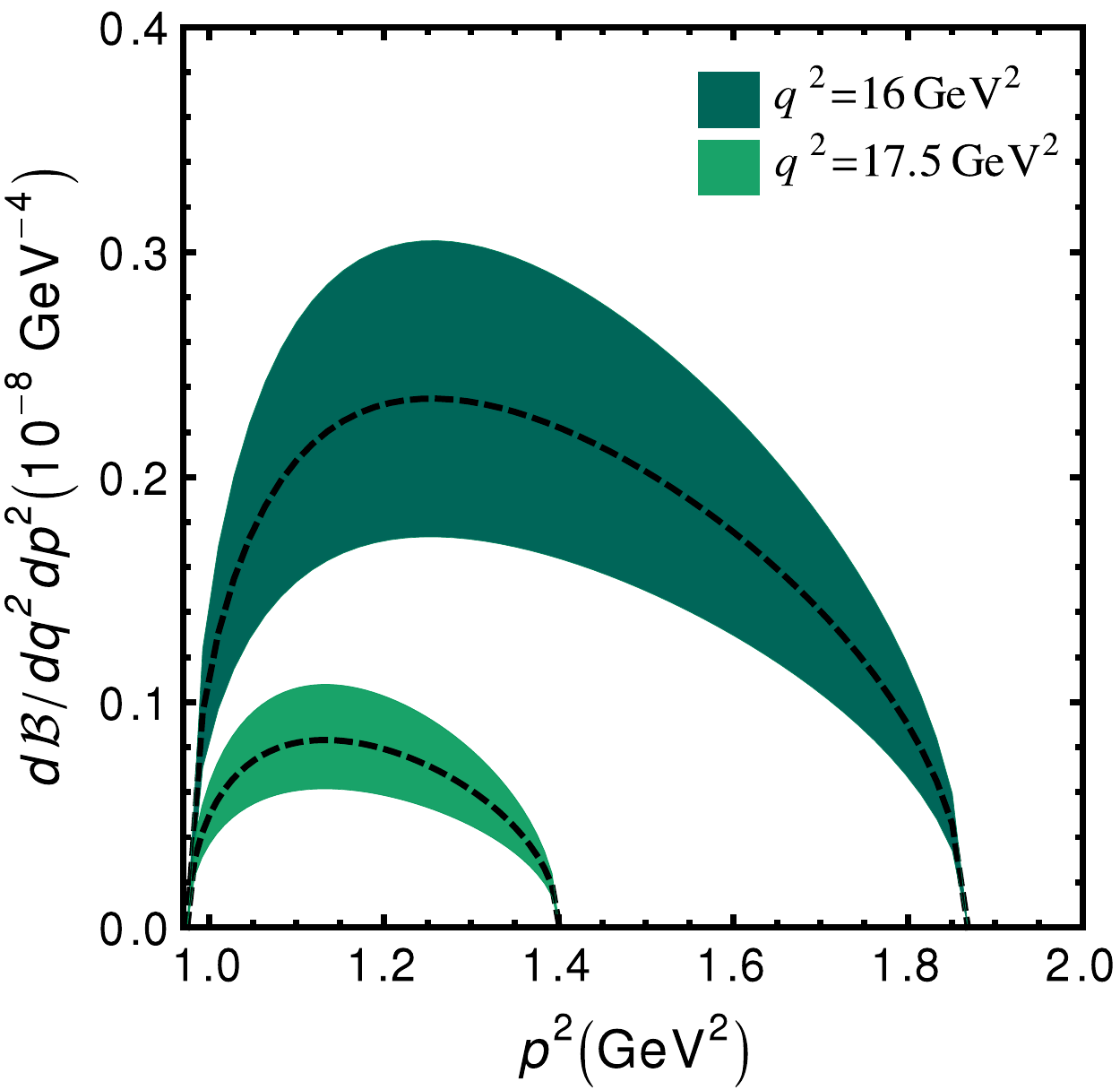}
}
\caption{\small Non-resonant $d{\cal{B}}(\bar B_s \to \bar K  K \ell  
\ell)/dq^2$ (left) without $p^2$-cuts, in the P-wave 'signal'  
window and the S+P-wave 'total' window, and
$d^2{\cal{B}}(\bar B_s \to \bar K K \ell \ell)/d q^2 dp^2$  (right)  
for fixed $q^2=16,17.5 \, \mbox{GeV}^2$ (outer and inner curve, respectively)
in the SM, see text for details. $\bar B_s \to \bar K K$ form factors are taken from HH$\chi$PT,  Eq.~(\ref{eq:ffinputBs}), and include parametric uncertainties only. Dashed lines are for central values of the input parameters.
\label{fig:BsKK}
}
\end{figure}

\begin{figure}[ht]
\centering{
\includegraphics[width=0.4\textwidth]{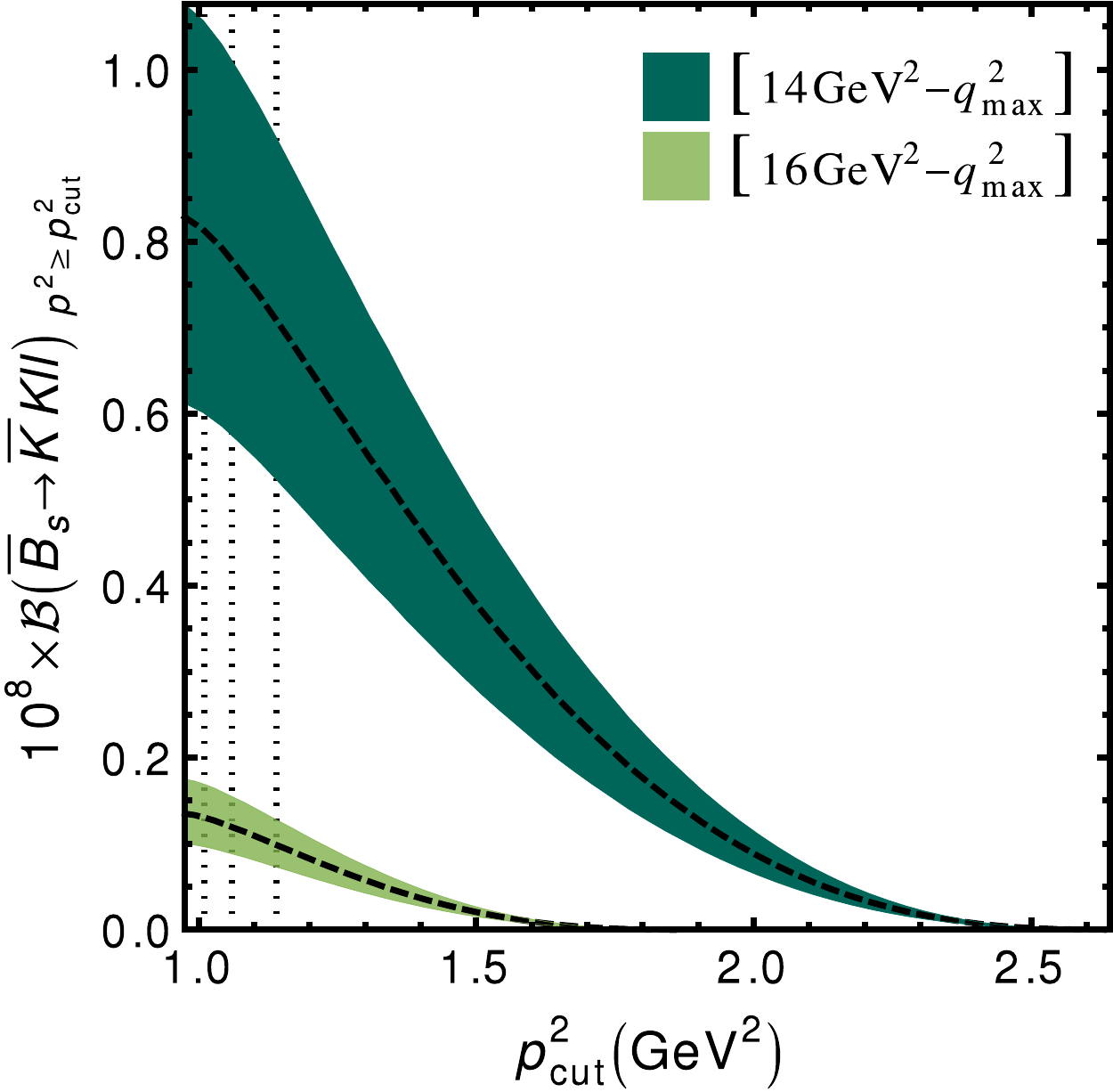}
}
\caption{\small The branching fraction of non-resonant $\bar B_s \to \bar K K \ell \ell$ decays in the SM, integrated over the low-recoil region from $(14,16) \, \mbox{GeV}^2-q^2_{\rm max}$ as a function of the lower $p^2$-integration boundary, analogously to the definition in Eq.~\eqref{eq:BRp2cut}. The vertical lines indicate (from left to right) the lower and upper bound for the P window, and the upper cut for the S+P window.
$\bar B_s \to \bar K K$ form factors are taken from HH$\chi$PT,  Eq.~(\ref{eq:ffinputBs}), and include parametric uncertainties only.
\label{fig:nr-cutBS}
}
\end{figure}

The differential branching fractions for non-resonant \BsKKll{} decays 
in the SM at low recoil  are shown in Fig.~\ref{fig:BsKK}.   
The SM branching ratio as a function of the low $p^2$-integration  
cut is presented in Fig.~\ref{fig:nr-cutBS}.
The form factors from HH$\chi$PT used in this prediction 
are given in Eq.~(\ref{eq:ffinputBs}).  Theory uncertainties   
as discussed in Section \ref{sec:kstar} apply likewise.
Due to the tighter P and S+P cuts the suppression of the non-resonant  
rates is by one order of magnitude more efficient
than for $\bar B \to \bar K \pi  \ell \ell$ decays.

Given the absence of additional interfering resonances, the neglect of doubly suppressed contributions, {\it i.e.} S-S,
S-D, and D-D interference, is clearly justified for \BsKKll, simplifying the angular analysis greatly.
Furthermore, the possibility to isolate the resonant contribution from $\bar B_s\to\phi\ell\ell$ so well offers
opportunities for using the non-resonant decays as a signal mode: outside the 
signal region for the $\phi$, they provide the dominant contribution. The null tests discussed for 
vanishing strong phase differences in \BKpill{} in
Section~\ref{sec:null} are actually probing new physics in the \BsKKll{} decay: a potential
significant measurement of one of the coefficients $J_{(8,9)x}$ in this decay would 
indicate new physics. $J_{7x}$, however, would remain zero to very good approximation.

\section{Conclusions \label{sec:conclusions}}

We present model-independent distributions for non-resonant $\bar B  
\to \bar K \pi \ell \ell$ and $\bar B_s \to \bar K  K  \ell \ell$ decays, where $ 
\ell=e,\mu$, at low hadronic recoil. To benefit from the OPE in $1/m_b$ we give  
improved Isgur-Wise form factor relations between the vector and  
tensor currents, as shown in Section~\ref{sec:FF}. These relations
follow from the equations of motion and make the universal structure  
of the helicity amplitudes inherited from the kinematic endpoint  
\cite{Hiller:2013cza} manifest. The exact endpoint  
relations for weak decays \cite{Hiller:2013cza} are extended to non-resonant  
decays in Section~\ref{sec:ep}. The only non-vanishing amplitudes  
at the kinematic endpoint in \BKpill{} decays come from P-wave states.
The behavior of the non-resonant decays at zero recoil therefore
reflects features of $\bar B \to \bar K^*(892) \ell \ell$ decays,  
however at different values of the dilepton invariant mass.
This is shown explicitly for the observable $F_L$.

The non-resonant modes constitute a background to precision tests of the SM with
 $\bar B \to  \bar K^* (\ \to \bar K \pi) \ell \ell$ and $\bar B_s  
\to  \phi (\to \bar K  K ) \ell \ell$ decays. While the  
branching fractions of non-resonant decays are at the level of $10^{-8}$ in the SM,
and hence only about an order of magnitude smaller than the P-wave  
signal modes, kinematic cuts suppress the non-resonant  rates in the 
P-wave analyses efficiently.
We also find that the branching fraction of resonant S-wave background is subdominant
to the branching fraction of the non-resonant decays.

Additional opportunities in \BsKKll{} follow from the fact that the $K\bar K$ distribution
is very simple and contains essentially only one pronounced resonance which is very narrow.
Once the $\phi$ is removed, the spectrum is given by the non-resonant decay which can be used
as an FCNC test.

Our numerical estimates are based on HH$\chi$PT $\bar B \to \bar K \pi$ form factors
which have been extrapolated in parts of the phase space beyond their nominal region of validity.
However, this affects the lower dilepton mass region more than the region closer to the endpoint.
Improvement of these approximations would require 
complementary and more precise information on the form factors available,  
for instance from lattice QCD \cite{Briceno:2014uqa}.

Angular analyses when non-resonant decays are included become significantly more  
involved as an infinite tower of states with different angular momenta is present.
The approximation of the non-resonant state to only the lowest S, P and D waves 
is sufficient to a percent-level precision in the rate. 
The full angular distribution in this approximation is given in Eq.~(\ref{eq:full}).
We summarize the qualitatively new ingredients in the study of non- 
resonant modes:
{\it i)}
Access to further combinations of Wilson coefficients  $\delta \rho$  
and $\rho_2^-$, as shown in Section \ref{sec:sd}, that are not present in $\bar B \to  
\bar K^{(*)} \ell \ell$ analyses,
{\it ii)} the possibility to probe strong phase differences using the  
interference with the resonant contributions,  
and {\it iii)}  new
contributions to null tests of $\bar B \to  
\bar K^{(*)} \ell \ell$ decays. 
The latter two items are discussed in Section~\ref{sec:angularpheno}.

Before closing we note that one may also consider lepton-universality breaking effects between dielectron and dimuon final states through the ratio 
$R_{\bar K \pi}$, see Eq.~(\ref{eq:RFpi}), and its $\bar B_s\to \bar K K\ell\ell$ counterpart. 
This is of interest in view of the recent preliminary data by LHCb on
the related ratio for $\bar B \to \bar K \ell \ell$, $R_K =0.745 \pm^{0.090}_{0.074} \pm 0.036$ \cite{Aaij:2014ora} in the bin $1 \, \mbox{GeV}^2 \leq q^2 < 6  \, \mbox{GeV}^2$.
Comparing to unity and adding systematic and statistical errors in quadrature,
this constitutes a $2.6 \sigma$ hint for lepton-flavor non-universal physics beyond the SM.
Interpreted within the SM+SM$^\prime$ basis, this yields (at $1 \sigma$)  roughly  
\begin{align} \label{eq:Xbounds}
-1.5 \lesssim {\rm Re}(X^\mu-X^e) \lesssim  -0.7, \quad X^\ell=C_9^{{\rm NP} \ell} +C_9^{\prime  \ell}-
(C_{10}^{{\rm NP}  \ell} +C_{10}^{\prime \ell}) \,  , \quad \ell=e, \mu \, ,
\end{align}
pointing without further correlations from data or model-constraints to new physics in either $b \to s ee$, $b\to s \mu \mu$, or both, the latter however not being universal.
The study of the impact of (\ref{eq:Xbounds}) on the non-resonant distributions presented here, in particular $R_{\bar K \pi}$, is interesting but  beyond the scope of this work. However,
the recent data emphasize once more  the great potential of rare semileptonic decays to probe weak scale physics in and beyond the SM.

In addition  the $\bar B \to \bar K \pi \ell \ell$ distribution contributes to the high $q^2$-tail of  $\bar B \to X_s \ell \ell$ decays where the latter  
cease to be inclusive \cite{Buchalla:1998mt}. While having smaller rates than $\bar B \to \bar K  \ell \ell$ the non-resonant modes have a richer angular structure and constitute the dominant contribution to the forward-backward asymmetry for $q^2 >(m_B -m_{K^*})^2$.

We expect that this work supports the exploration of flavored  
processes and look forward to future analyses.

\begin{acknowledgments}
We are grateful to Damir Becirevic, Christoph Bobeth, Gerhard Buchalla, Ulf Mei\ss{}ner, Kostas Petridis, Danny van Dyk, and Roman Zwicky  for useful communication.
Diganta Das is supported by the DFG Research Unit FOR 1873 \lq\lq{}Quark Flavour Physics and Effective Field Theories\rq\rq{}.
The work by Martin Jung is supported in part by the Bundesministerium f\"ur Bildung und Forschung (BMBF).
Alex Shires is supported by the Emmy Noether Programme, grant number AL 1639/1-1.
Gudrun Hiller gratefully acknowledges the hospitality and stimulating atmosphere provided by the
Aspen Center for Physics during the final phase of this work.
\end{acknowledgments}


\appendix

\section{Parametric Input \label{app:input}}

 \begin{table}[ht]
\begin{center}
\begin{tabular}{c|c|c}
\hline\hline
Parameter           &     Value & Source \\\hline\hline
$|V_{ts}^* V_{tb}|$ & $0.0407 \pm 0.0011$ &  \cite{Charles:2004jd}\\
$\Gamma(B_0)$ & $(4.333 \pm 0.020) \cdot 10^{-13}$ GeV & \cite{PDG} \\
$\Gamma(B_s)$ & $(4.342 \pm 0.032) \cdot 10^{-13}$ GeV  & \cite{PDG} \\
$f_\pi $ & $130.4 \pm 0.2$ MeV & \cite{PDG} \\
$f_K $ & $156.2 \pm 0.7$ MeV &  \cite{PDG}$^\dagger$ \\
$f_{B_d} $ & $188 \pm 4$ MeV &  \cite{Dowdall:2013tga}\\
 $f_{B_s} $ & $224\pm 5$ MeV &  \cite{Dowdall:2013tga}\\
$g$ & $0.569 \pm 0.076$ & \cite{Flynn:2013kwa}$^\dagger$
\\\hline\hline
\end{tabular}
\end{center}
\caption{\label{tab:input}
Numerical input used in this work. $\Gamma(B_{0,s})$ denotes the mean total width. $^\dagger$Uncertainties added in quadrature.}
\end{table}

The lattice value of $g$  \cite{Flynn:2013kwa} obtained in the $B$ system is in good agreement with the one extracted from the $D^{*+} \to D^0 \pi^+$ decay rate,
$g=0.59 \pm 0.07$  \cite{PDG}. The main uncertainty on the latter stems from the $D^{*+}$ total width \cite{PDG}. The values should agree due to heavy quark symmetry.
We choose to use for the decay constants of the pseudoscalar mesons in the $SU(3)$ limit, $f^2$ and $f_B$, the values $f_\pi f_K$ and $f_{B_d}$, respectively. Note that $f_{B_d}/f_\pi\approx f_{B_s}/f_K$, as can be inferred from Table~\ref{tab:input}, albeit these corrections are beyond the scope of the HH$\chi$PT calculation employed in this work.

\section{Kinematics \label{app:kinematics}}

We consider the decay $\bar B^0\to \bar K\pi\ell^+\ell^-$ and define
\begin{align}
q&=p_{\ell^-}+p_{\ell^+}\,, &Q&=p_{\ell^-}-p_{\ell^+}\,,\\
p&=p_K+p_\pi=p_B-q\,, &P&=p_K-p_\pi\,.
\end{align}
This way, $q^2$ and $p^2$ denote the invariant mass squared of the dilepton- and $\bar K\pi$-system, respectively.
Assuming $m_\ell=0$,
the relevant scalar products read 
\begin{eqnarray}
p_B\cdot p_K &=&
\frac{1}{4p^2}\left[(m_B^2-q^2+p^2)(p^2+m_K^2-m_\pi^2)-\sqrt{ \lambda  \lambda_{p}}\cos\theta_K\right]\,,\\
p_B\cdot p_\pi &=&\frac{1}{4p^2}\left[(m_B^2-q^2+p^2)(p^2-m_K^2+m_\pi^2)+\sqrt{ \lambda  \lambda_{p}} \cos\theta_K\right]\,,\\
p_B\cdot p_{\ell^-} &=& \frac{1}{4}\left[(m_B^2+q^2-p^2)-\lambda^{1/2}\cos\theta_\ell\right]\,,\\
p_B\cdot p_{\ell^+} &=& \frac{1}{4}\left[(m_B^2+q^2-p^2)+\lambda^{1/2}\cos\theta_\ell\right]\,,\\
p_{\ell^-}\cdot p_{\ell^+} &=& \frac{q^2}{2}\,,\\
p_K\cdot p_\pi &=& \frac{p^2-m_K^2-m_\pi^2}{2}\,,\\[2ex]
P\cdot p &=& m_K^2-m_\pi^2\,,\\
Q\cdot q &=& 0\,,\\
p\cdot q &=& \frac{m_B^2-q^2-p^2}{2}\,,\\
p\cdot Q &=& -\frac{1}{2} \sqrt{\lambda} \cos\theta_\ell\,,\\
P\cdot q &=& \frac{1}{2p^2}\left[(m_B^2-p^2-q^2)(m_K^2-m_\pi^2)-\sqrt{ \lambda  \lambda_{p}} \cos\theta_K\right]\,,\\
P\cdot Q &=& \frac{1}{2p^2}\left[(m_B^2-q^2-p^2)\lambda^{1/2}_{p}\cos\theta_K\cos\theta_\ell-(m_K^2-m_\pi^2)\lambda^{1/2} \cos\theta_\ell\right.\nonumber\\
&&\left.\,-2\sqrt{q^2 p^2 \lambda_{p}}\cos\phi\sin\theta_K\sin\theta_\ell\right]\,,\\
\epsilon_{\mu\nu\rho\sigma}P^\mu p^\nu Q^\rho q^\sigma&=&\frac{1}{2} \sqrt{ \lambda  \lambda_{p} \frac{q^2}{p^2}}\sin\phi\sin\theta_K\sin\theta_\ell\,, \\
p^2+P^2&=&2(m_K^2+m_\pi^2)\,, \quad  Q^2=-q^2 \, ,
\end{eqnarray}
where the phase space factors $\lambda$ and $ \lambda_p$ are defined after Eq.~(\ref{eq:Nnr}).

These expressions confirm the ones given by Wise et al. \cite{Lee:1992ih} ($LLW$) when taking into account the following differences ($DHJS$: this work):
\begin{align}
 \epsilon_{0123}^{LLW}=1=- \epsilon_{0123}^{DHJS} ,\,\,
 \theta_K^{LLW}=\pi-\theta_K^{DHJS} ,\,\, \phi^{LLW}=(\pi+\phi^{DHJS})\!\!\!\mod(2\pi)\,.
\end{align}
Although stated differently, the definitions for $\theta_\ell$ agree.

\section{Five-fold differential rate}
The four-body phase space can be calculated by considering successive two-body transitions with the momenta $p_B\to p(\to p_\pi p_K)q(\to p_{\ell^+}p_{\ell^-})$,
yielding (see, \emph{e.g.}, \cite{Widhalmthesis})
\begin{equation}
 dR_4^{\text{LIPS}}=\frac{1}{4(4\pi)^6}\frac{\lambda^{1/2}\lambda^{1/2}_p}{p^2\,m_B^2}
 dp^2 dq^2
 d\text{cos}\theta_K d\text{cos}\theta_\ell d\phi \, .
\end{equation}
Thanks to the proportionality of the two matrix elements in Eqs.~\eqref{eq::FFLLW}, \eqref{eq::FFLLW2}, that is, the Isgur-Wise relations given in Eq.~\eqref{eq:iwr}, the fully differential angular distribution for the SM basis can be expressed as
\begin{eqnarray}
d^5\Gamma &=& \frac{1}{2m_B}\sum_{\rm{spins}}|\mathcal M|^2 dR_4^{\rm{LIPS}}\\
&=& \frac{1}{2m_B}\frac{\alpha^2_eG_F^2|V_{tb}V_{ts}^*|^2}{2\pi^2}H_{\mu\nu}^{ij}\left[\rho_1 L_S^{\mu\nu}-2\rho_2 L_A^{\mu\nu}\right]dR_4^{\rm{LIPS}}\\
&=&\frac{\mathcal N^2_{nr}}{2\pi}H_{\mu\nu}^{ij}\left[\rho_1 L_S^{\mu\nu}-2\rho_2 L_A^{\mu\nu}\right]dq^2dp^2d\cos\theta_Kd\cos\theta_\ell d\phi\,,
\end{eqnarray}
with the short-distance couplings $\rho_{1,2}$ defined in Eq.~\eqref{eq:SMrho}.
This shows again that at low recoil in the SM basis any observable can be sensitive to two combinations of Wilson coefficients, only; including the primed operators then leads to the additional combinations given in Eq.~\eqref{eq:SD}.
Since the form factors depend on $p^2,q^2,\theta_K$, only, we will group the squared matrix element as follows, following \cite{Lee:1992ih}:
\begin{eqnarray}
\mathcal N_{nr}^2H_{\mu\nu}^{ij}L^{\mu\nu} &=& |c_{ij}|^2\left[\rho_1\!\!\!\!\!\sum_{i=1-4,8,9}\!\!\!\!\!c_i(\theta_\ell,\phi) I_i(q^2,p^2,\theta_K)-2\rho_2\sum_{i=5}^7c_i(\theta_\ell,\phi) I_i(q^2,p^2,\theta_K)\right]\,,\label{eq::Iamps}
\end{eqnarray}
with the coefficients  $c_i$ given in Eq.~(\ref{eq:ci}).

For the explicit calculation, we start with the 
leptonic tensor\footnote{The expressions in \cite{Lee:1992ih} correspond to $C_9=+1$, $C_{10}=-1$.}:
\begin{eqnarray}
L^{\mu\nu}&=&\rho_1L^{\mu\nu}_S-2\rho_2L^{\mu\nu}_A\,,\\
L^{\mu\nu}_S &=&\frac{1}{2}\left[q^\mu q^\nu-Q^\mu Q^\nu-q^2g^{\mu\nu}\right]\,,\\
L^{\mu\nu}_A &=&-\frac{i}{2}\epsilon^{\alpha\mu\gamma\nu}q_\alpha Q_\gamma\,.
\end{eqnarray}
The hadronic tensor is given as
\begin{eqnarray}
H_{\mu\nu}^{ij} &=& \langle \bar K^i(p_K) \pi^j(p_\pi)|\bar s\gamma_\mu(1-\gamma_5)b|\bar B(p_B)\rangle\langle \bar K^i(p_K) \pi^j(p_\pi)|\bar s\gamma_\nu(1-\gamma_5)b|\bar B(p_B)\rangle^*\,,
\end{eqnarray}
where we parametrize the matrix element as in Eq.~\eqref{eq::FFLLW}.
Clearly, the two contributing terms are
\begin{eqnarray}
L^{\mu\nu}H_{\mu\nu}^{ij} &=& \rho_1L^{\mu\nu}_S H_{\mu\nu,S}^{ij}-2\rho_2L^{\mu\nu}_A H_{\mu\nu,A}^{ij}\,.
\end{eqnarray}
We obtain 
\begin{align}
L^{\mu\nu}_S H_{\mu\nu,S}^{ij} =& |c^{ij}|^2L^{\mu\nu}_S\{ {\rm Re}\left[(w_+p_\mu+w_-P_\mu +q_\mu r)(w_+p_\nu+w_-P_\nu +q_\nu r)^*\right] \phantom{P^\beta_\nu}\right.\nonumber\\
&+|h|^2\epsilon_{\mu\alpha\beta\gamma}\epsilon_{\nu\rho\sigma\tau}p_B^\alpha p^\beta P^\gamma p_B^\rho p^\sigma P^\tau\nonumber\\
&\left.+2{\rm Im}\left[h^*(w_+p_\nu+w_-P_\nu +q_\nu r)\right]\epsilon_{\mu\alpha\beta\gamma}p_B^\alpha p^\beta P^\gamma\}\,,\\
L^{\mu\nu}_A H_{\mu\nu,A}^{ij} =& i|c^{ij}|^2L^{\mu\nu}_A\{ {\rm Im}[(w_+p_\mu+w_-P_\mu +q_\mu r)(w_+p_\nu+w_-P_\nu +q_\nu r)^*] \phantom{P^\beta_\nu}\right.\nonumber\\
&\left.+2{\rm Re}(h^*(w_+p_\nu+w_-P_\nu +q_\nu r)\epsilon_{\mu\alpha\beta\gamma}p_B^\alpha p^\beta P^\gamma\}\,,
\end{align}
the calculation of which is lengthy, but straight-forward. Expressing the two contributions in terms of the Lorentz-invariants calculated before, we get ($y=w_+ p+w_-P$)
\begin{eqnarray}
L_S^{\mu\nu}H_{\mu\nu,S}^{ij} &=& \frac{|c^{ij}|^2}{2} \{|w_+|^2\left[(p\cdot q)^2-(p\cdot Q)^2-q^2p^2\right]+|w_-|^2\left[(P\cdot q)^2-(P\cdot Q)^2-q^2P^2\right]\right.\nonumber\\
&&+2{\rm Re}(w_+w_-^*)(p\cdot q P\cdot q-p\cdot QP\cdot Q-q^2p\cdot P)-|h|^2(\epsilon_{\mu\nu\rho\sigma} P^\mu p^\nu Q^\rho q^\sigma)^2\nonumber\\
&&+|h|^2q^2\left[q^2\left(p^2P^2-(p\cdot P)^2\right)-p\cdot q(p\cdot q P^2-p\cdot P q\cdot P)+q\cdot P(p\cdot q p\cdot P-p^2q\cdot P)\right]\nonumber\\
&&\left.+\epsilon_{\mu\nu\rho\sigma} P^\mu p^\nu Q^\rho q^\sigma\left[{\rm Im}(w_+h^*)p\cdot Q+{\rm Im}(w_-h^*)P\cdot Q\right]\} \, ,\\
L_A^{\mu\nu}H_{\mu\nu,A}^{ij} &=& |c^{ij}|^2 \{
	{\rm Re}\left[h^*q^2(p\cdot Q P\cdot y-p\cdot yP\cdot Q)+h^*q\cdot y(p\cdot q P\cdot Q-p\cdot Q q\cdot P)\right]\right.\nonumber\\
&&\left.-\epsilon_{\mu\nu\rho\sigma} P^\mu p^\nu Q^\rho q^\sigma {\rm Im}(w_+w_-^*)
	\} \, . 
\end{eqnarray}

\section{HH$\boldsymbol{\chi}$PT-non-resonant  form factors \label{app:nonresonant}}
The $\bar B \to \bar K \pi$ matrix element can be parameterized as, following Ref.~\cite{Buchalla:1998mt}\footnote{We remove the $1/q^2$ employed in  Ref.~\cite{Buchalla:1998mt}  and add a factor $m_B$ for dimensional reasons in the definition of the tensor matrix element.},
\begin{align}
\langle\bar K^i(p_K)\pi^j(p_\pi)|\bar s\gamma_\mu(1-\gamma_5)b|
\bar B(p_B)\rangle& =
i c_{ij} \left[ a p_{\pi \mu} + b p_{K \mu} + c p_{B \mu}
       -2 i h \varepsilon_{\mu\alpha\beta\gamma}p_B^\alpha p_K^\beta
       p_\pi^\gamma \right]  \, , \\
       \langle\bar K^i(p_K)\pi^j(p_\pi)|
\bar s i q^\nu \sigma_{\mu \nu}(1+\gamma_5)   b|\bar B(p_B)\rangle &=
- i c_{ij} m_B \left[ a' p_{\pi \mu} + b' p_{K \mu} + c' p_{B \mu}
       -2 i h' \varepsilon_{\mu\alpha\beta\gamma}p_B^\alpha p_K^\beta
       p_\pi^\gamma \right]  \, , 
\end{align}
with form factors $a^{\scriptscriptstyle{(\prime)}}, b^{\scriptscriptstyle{(\prime)}}, c^{\scriptscriptstyle{(\prime)}}, h^{\scriptscriptstyle{(\prime)}}$, which depend on $q^2,p^2$ and $\cos \theta_K$.
To lowest order HH$\chi$PT~\cite{Buchalla:1998mt} they read
\begin{align}
a &= \frac{gf_B}{f^2} \frac{m_B}{v\cdot p_\pi+\Delta}~, 
\qquad\qquad b=0~,\\
c &= \frac{f_B}{2f^2}\left[1
-2g\frac{v\cdot p_\pi}{v\cdot p_\pi+\Delta}
-\frac{v\cdot (p_K-p_\pi)}{v\cdot (p_K+p_\pi)+\mu_s}  -2g^2 \frac{p_K\cdot p_\pi -v\cdot p_K v\cdot p_\pi}
{[v\cdot p_\pi+\Delta][v\cdot (p_K+p_\pi)+\mu_s]} \right],\\
\label{hchpt}
h &= \frac{g^2 f_B}{2f^2} \frac{1}{ 
[v\cdot p_\pi+\Delta][v\cdot (p_K+p_\pi)+\Delta+\mu_s]} \, , \\
a' &= \frac{g f_B}{f^2 (v\cdot p_\pi+\Delta)}
 \Biggl[m_B-v\cdot p_K-v\cdot p_\pi 
+g\frac{v\cdot p_K v\cdot( p_K+p_\pi)-
  p_K\cdot p_\pi-m^2_K}{v\cdot(p_K +p_\pi)+\Delta+\mu_s}\Biggr]~,\\
b' &= \frac{g^2 f_B }{f^2 (v\cdot p_\pi+\Delta)}
 \frac{p_K\cdot p_\pi+m^2_\pi-v\cdot p_\pi\ v\cdot(p_K+p_\pi)}{
       v\cdot(p_K +p_\pi)+\Delta+\mu_s}~,\\
c' &= -\frac{g f_B}{f^2 m_B (v\cdot p_\pi+\Delta)}
 \Biggl[m_B v\cdot p_\pi-m^2_\pi-p_K\cdot p_\pi +g\frac{p_K\cdot p_\pi\ v\cdot(p_K -p_\pi)-m^2_K v\cdot p_\pi+
  m^2_\pi v\cdot p_K}{v\cdot(p_K +p_\pi)+\Delta+\mu_s}\Biggr]~,\\
h' &= \frac{g f_B}{2 f^2 m_B (v\cdot p_\pi+\Delta)}
 \left[1+g\frac{m_B-v\cdot p_K - 
      v\cdot p_\pi}{v\cdot(p_K +p_\pi)+\Delta+\mu_s}\right]~.
\end{align}

 Keeping leading terms in the expressions for the primed form factors   only (with $g \sim 1$), one obtains
 \begin{align}\label{eq::IIW}
 a^{(\prime)}=\frac{gf_B m_B}{f^2 (v\cdot p_\pi+\Delta)}, \quad b^{(\prime)}=0 , \quad c^\prime=-\frac{g f_B v \cdot p_\pi}{f^2  (v\cdot p_\pi+\Delta)}, \quad
 h^{(\prime)}=\frac{g^2 f_B}{2f^2 (v\cdot p_\pi+\Delta)} \frac{1}{ v\cdot p+\Delta+\mu_s} \, .
 \end{align}
 Note that $c^{(\prime)}/a = {\cal{O}}(1/m_b)$ holds.

\section{S-, P-, and D-wave contributions \label{app:resonant}}

{}From the full angular distribution, given in Eq.~(\ref{eq:full}), one can read off its contributions from the S-, P-, and D-waves and their interference. At low recoil, the angular coefficients $J_{ix}$ can be expressed in terms of short-distance couplings, presented in Section \ref{sec:sd},
and form factors $F_{i \ell}=F_{i \ell}(q^2,p^2)$, $i=0, \parallel, \perp $. The latter receive contributions from the angular expansion of the non-resonant amplitudes, Eq.~(\ref{eq:expand}), and
from decays of spin 0,1,2 resonances as given below in Section~\ref{sec:012}. Explicitly, the coefficients read as follows:
\begin{align}
J_{1cc} & = \frac{1}{8} \left[|F_{0S}|^2+ |F_{0P}|^2+ |F_{0D}|^2+ 2 {\rm Re}(F_{0S} F_{0D}^*)\right] \rho_1^-  \, ,\\
J_{1ss} & = \frac{1}{8} \left[ \left(|F_{0S}|^2+ \frac{1}{4} |F_{0D}|^2-  {\rm Re}(F_{0S} F_{0D}^*) 
+\frac{3}{2}  |F_{\parallel P}|^2\right) \rho_1^- + \frac{3}{2} |F_{\perp P}|^2 \rho_1^+ \right] \, ,\\
J_{1c} & =\frac{1}{4}  \left[{\rm Re}(F_{0P} F_{0S}^*)+ {\rm Re}(F_{0P} F_{0D}^*)\right] \rho_1^-  \, ,\\
J_{1ssc}&= \frac{3}{8} \left[  \left(-{\rm Re}(F_{0P} F_{0D}^*)+3 {\rm Re}(F_{\parallel P} F_{\parallel D}^*)\right) \rho_1^-+ 3 {\rm Re}(F_{\perp P} F_{\perp D}^*) \rho_1^+ \right] \, , \\
J_{1sscc}&=  \frac{9}{16} \left[  \left(- \frac{1}{2} |F_{0D}|^2 + 3 |F_{\parallel D}|^2\right) \rho_1^-+   3 |F_{\perp D}|^2 \rho_1^+ \right]  \, ,\\
J_{2cc} & =- J_{1cc} \, ,\\
J_{2ss} & = -\frac{1}{8} \left[ \left(|F_{0S}|^2+ \frac{1}{4} |F_{0D}|^2-  {\rm Re}(F_{0S} F_{0D}^*)
-\frac{1}{2}  |F_{\parallel P}|^2\right) \rho_1^- - \frac{1}{2} |F_{\perp P}|^2 \rho_1^+ \right] \, ,\ \\
J_{2c} & =-J_{1c} \, ,\\
J_{2ssc}&= \frac{3}{8} \left[  \left({\rm Re}(F_{0P} F_{0D}^*)+ {\rm Re}(F_{\parallel P} F_{\parallel D}^*)\right) \rho_1^-+  {\rm Re} (F_{\perp P} F_{\perp D}^*) \rho_1^+ \right]  \, , \\
J_{2sscc}&=  \frac{9}{16} \left[  \left( \frac{1}{2} |F_{0D}|^2 +  |F_{\parallel D}|^2\right) \rho_1^-+    |F_{\perp D}|^2 \rho_1^+ \right]  \, , \\
J_{3} & =\frac{1}{8}\left( |F_{\perp P}|^2 \rho_1^+ -  |F_{\parallel P}|^2 \rho_1^-\right)  \, ,  \\
J_{3cc}& = \frac{9}{8}\left( |F_{\perp D}|^2 \rho_1^+ -  |F_{\parallel D}|^2 \rho_1^-\right) \, ,  \\
J_{3c} & =\frac{3}{4}\left( {\rm Re} (F_{\perp P} F_{\perp D}^*) \,   \rho_1^+ -  {\rm Re}(F_{\parallel P}  F_{\parallel D}^*)  \, \rho_1^-\right) \, ,   \\
J_{4cc} & = \frac{1}{4}{\rm Re} \left(F_{0S} F_{\parallel P}^* + 3 F_{0P} F_{\parallel D}^* + 
F_{0D} F_{\parallel P}^*\right)  \rho_1^- \, ,  \\
J_{4ss} &= \frac{1}{4}{\rm Re} \left(F_{0S} F_{\parallel P}^* -\frac{1}{2}  F_{0D} F_{\parallel P}^*\right)  \rho_1^- \, ,  \\
J_{4c} &= \frac{1}{4}{\rm Re} \left(F_{0P} F_{\parallel P}^* + 3 F_{0S} F_{\parallel D}^* + 3 F_{0D} F_{\parallel D}^*\right)  \rho_1^- \, ,  \\
J_{4ssc} &= -\frac{9}{8} {\rm Re}  (F_{0D} F_{\parallel D}^*)  \rho_1^- \, ,  \\
J_{5cc}&=-{\rm Re}\left(3F_{0P}F_{\perp D}^*+F_{0D}F_{\perp P}^*+F_{0S}F_{\perp P}^*\right){\rm Re}\rho_2^+
        -{\rm Im}\left(3F_{0P}F_{\perp D}^*+F_{0D}F_{\perp P}^*+F_{0S}F_{\perp P}^*\right){\rm Im}\rho_2^- \, ,  \\
J_{5ss}&=\frac{1}{2}{\rm Re}\left(F_{0D}F_{\perp P}^*-2F_{0S}F_{\perp P}^*\right){\rm Re}\rho_2^+
        +\frac{1}{2}{\rm Im}\left(F_{0D}F_{\perp P}^*-2F_{0S}F_{\perp P}^*\right){\rm Im}\rho_2^- \, ,  \\
J_{5c}&=-{\rm Re}\left(3F_{0D}F_{\perp D}^*+3F_{0S}F_{\perp D}^*+F_{0P}F_{\perp P}^*\right){\rm Re}\rho_2^+
        -{\rm Im}\left(3F_{0D}F_{\perp D}^*+3F_{0S}F_{\perp D}^*+F_{0P}F_{\perp P}^*\right){\rm Im}\rho_2^- \, ,  \\
J_{5ssc}&=\frac{9}{2}{\rm Re}(F_{0D}F_{\perp D}^*){\rm Re}\rho_2^+
        +\frac{9}{2}{\rm Im}(F_{0D}F_{\perp D}^*){\rm Im}\rho_2^-\, ,  \\
J_{6cc}&=-9{\rm Re}\left(F_{\|D}F_{\perp D}^*\right){\rm Re}\rho_2^+-9{\rm Im}\left(F_{\|D}F_{\perp D}^*\right){\rm Im}\rho_2^- \, ,  \\
J_{6}&=-{\rm Re}\left(F_{\|P} F_{\perp P}^*\right){\rm Re}\rho_2^+-{\rm Im}\left(F_{\|P} F_{\perp P}^*\right){\rm Im}\rho_2^- \, ,  \\
J_{6c}&=-3{\rm Re}\left(F_{\|P}F_{\perp D}^*+F_{\|D}F_{\perp P}^*\right){\rm Re}\rho_2^+
         -3{\rm Im}\left(F_{\|P}F_{\perp D}^*+F_{\|D}F_{\perp P}^*\right){\rm Im}\rho_2^- \, ,  \\
J_{7cc}&=-{\rm Im}\left(3F_{0 P}F_{\| D}^*+F_{0 D}F_{\| P}^*+F_{0S}F_{\| P}^*\right)\delta\rho \, ,  \\
J_{7ss}&=\frac{1}{2} {\rm Im}\left(F_{0 D}F_{\| P}^*-2F_{0 S}F_{\| P}^*\right)\delta\rho \, ,  \\
J_{7c}&=- {\rm Im}\left(3F_{0 D}F_{\| D}^*+3F_{0 S}F_{\| D}^*+F_{0 P}F_{\| P}^*\right)\delta\rho \, ,  \\
J_{7ssc}&=\frac{9}{2} {\rm Im}\left(F_{0 D}F_{\| D}^*\right)\delta\rho \, ,  \\
J_{8cc}&=-\frac{1}{2}{\rm Re}\left(3F_{0P}F_{\perp D}^*+F_{0D}F_{\perp P}^*+F_{0S}F_{\perp P}^*\right){\rm Im}\rho_2^+
        +\frac{1}{2}{\rm Im}\left(3F_{0P}F_{\perp D}^*+F_{0D}F_{\perp P}^*+F_{0S}F_{\perp P}^*\right){\rm Re}\rho_2^-\, ,  \\
J_{8ss}&=\frac{1}{4}{\rm Re}\left(F_{0D}F_{\perp P}^*-2F_{0S}F_{\perp P}^*\right){\rm Im}\rho_2^+
        -\frac{1}{4}{\rm Im}\left(F_{0D}F_{\perp P}^*-2F_{0S}F_{\perp P}^*\right){\rm Re}\rho_2^-\, ,  \\
J_{8c}&=-\frac{1}{2}{\rm Re}\left(3F_{0D}F_{\perp D}^*+3F_{0S}F_{\perp D}^*+F_{0P}F_{\perp P}^*\right){\rm Im}\rho_2^+
        +\frac{1}{2}{\rm Im}\left(3F_{0D}F_{\perp D}^*+3F_{0S}F_{\perp D}^*+F_{0P}F_{\perp P}^*\right){\rm Re}\rho_2^- \, ,  \\
J_{8ssc}&=\frac{9}{4}{\rm Re}\left(F_{0D}F_{\perp D}^*\right){\rm Im}\rho_2^+
        -\frac{9}{4}{\rm Im}\left(F_{0D}F_{\perp D}^*\right){\rm Re}\rho_2^- \, ,  \\
J_{9cc}&=\frac{9}{2}{\rm Re}\left(F_{\perp D}F_{\| D}^*\right){\rm Im}\rho_2^++\frac{9}{2}{\rm Im}\left(F_{\perp D}F_{\| D}^*\right){\rm Re}\rho_2^- \, ,  \\
J_{9}&=\frac{1}{2}{\rm Re}\left(F_{\perp P} F_{\| P}^*\right){\rm Im}\rho_2^++\frac{1}{2}{\rm Im}\left(F_{\perp P} F_{\| P}^*\right){\rm Re}\rho_2^- \, ,  \\
J_{9c}&=\frac{3}{2}{\rm Re}\left(F_{\perp P}F_{\| D}^*+F_{\perp D}F_{\| P}^*\right){\rm Im}\rho_2^+
         +\frac{3}{2}{\rm Im}\left(F_{\perp P}F_{\| D}^*+F_{\perp D}F_{\| P}^*\right){\rm Re}\rho_2^- \, .
\end{align}

\subsection{Identifying resonant kaon contributions with Spin 0,1 and 2 \label{sec:012}}

The pure resonant S,P-contributions can be taken in the full dimension 6 operator basis from \cite{Bobeth:2012vn}, the $D$-wave contribution from \cite{Li:2010ra}. Based on these expressions, one can identify the S-, P-, and D-wave contributions in zero width approximation up to a strong phase $\delta_J$ as follows:

The S-wave decay rate can be written as
\begin{align}
\frac{d^2 \Gamma (S)}{d q^2  d \cos \theta_K }  &= \int\! d p^2\, \delta\! \left(p^2-m_{K_0^*}^2\right) \left.\frac{d^3 \Gamma }{d q^2 d p^2 d \cos \theta_K } \right|_S =  \frac{ \rho_1^-}{3} |F_{0S}|^2\,, \\
\frac{d  \Gamma (S)}{d q^2 } &  =  \frac{2 \rho_1^-}{3}  |F_{0S}|^2\,.
\end{align}
Comparison with the standard form, which can be extracted \emph{e.g.} from Eq.~(65) in \cite{Bobeth:2012vn}, yields
\begin{align}
F_{0S}  &= \tilde f_+ \sqrt{\frac{\Gamma_0}{2}}\, \lambda_0^{3/4}\, e^{i \delta_S}\,,
\end{align}
where the $ \bar B \to$ scalar form factor $\tilde f_+(q^2)$ is  defined as
\begin{eqnarray}
\langle\bar S(p)|\bar s\gamma^\mu \gamma_5 b|\bar B(p_B)\rangle
&=& \tilde f_+(q^2)\, (p_B+p)^\mu +q^\mu  (\ldots) \, .
\end{eqnarray}
Estimates for $\tilde f_+(q^2)$ exist for $K_0^*(1430)$ in QCDSR~\cite{Aliev:2007rq}, the lightfront quark model~\cite{Chen:2007na} and pQCD~\cite{Li:2008tk}, where also the $\kappa(800)$ form factor has been estimated.
All of these methods work for
large momentum transfer and their results have to be extrapolated to the small recoil region.
The normalization $\Gamma_0$ is given by
\begin{align} \label{eq:normK0}
  \Gamma_0 &
    = \frac{G_F^2 \alpha_e^2 |V_{tb}  V^*_{ts}|^2}{2^9 \pi^5 m_B^3}\,, &
  \lambda_0 \equiv \lambda(m_B^2, m_{K_0^*}^2, q^2)\,.
  \end{align}

The pure P-wave contribution to the doubly differential rate in zero-width approximation is given as 
\begin{align}
\frac{d^2 \Gamma (P)}{d q^2  d \cos \theta_K }  &= \int\!d p^2\, \delta\! \left(p^2-m_{K^*}^2\right) \left.\frac{d^3 \Gamma}{d q^2 d p^2 d \cos \theta_K }\right|_P =  
\frac{1}{3} \left[ \cos^2 \theta_K \rho_1^-|F_{0P}|^2 + \sin^2 \theta_K \left(\rho_1^+|F_{\perp P}|^2+ \rho_1^-|F_{\parallel P}|^2\right)\right]\,, \\
\frac{d  \Gamma (P)}{d q^2 } &  =  \frac{2}{9} \left[\rho_1^-\, |F_{0P}|^2 + 2 \left(\rho_1^+\,|F_{\perp P}|^2+ \rho_1^-\,|F_{\parallel P}|^2\right)\right]\,.
\end{align}
Finite-width effects can be easily included by replacing the $\delta$ distribution by the corresponding distribution.
Matching onto the standard form, as, {\it e.g.,} given in~\cite{Bobeth:2010wg}, yields
\begin{align}
F_{0P} =&  - 3 f_0  \, e^{i \delta_P}\, , \quad F_{\parallel P} =  - 3  \sqrt{ \frac{ 1}{ 2}}  \, f_\parallel \, e^{i \delta_P}\, , \quad 
F_{\perp P} =  3 \sqrt{ \frac{ 1}{ 2}} \, f_\perp  \, e^{i \delta_P}\, ,
\end{align}
where the relative signs are from matching onto the angular coefficients.

The requisite $ \bar B \to$ vector transversity form factors are defined as
\begin{equation}
\begin{aligned}
  f_{\perp} & ={\cal N }_{K^*} \frac{\sqrt{2\, \lambda_{K^*}}}{m_B + m_{K^*}} V\,, 
\\
  f_{\parallel}  & = {\cal N}_{K^*} \sqrt{2}\, (m_B + m_{K^*})\, A_1\,,
\\
  f_{0} & ={\cal N}_{K^*} \frac{(m_B^2 - m_{K^*}^2 - q^2) (m_B + m_{K^*})^2 A_1
   - \lambda_{K^*}\, A_2}{2\, m_{K^*} (m_B + m_{K^*}) \sqrt{q^2}}\,, 
\end{aligned}
\label{eq:Kstff}
\end{equation}
where $\lambda_{K^*} \equiv \lambda(m_B^2, m_{K^*}^2, q^2)$, the normalization factor is
\begin{align}
  \label{eq:NKstar}
  {\cal N}_{K^*} & = G_F   V_{tb}^{}V_{ts}^{*} \alpha_e\,
    \sqrt{\frac{q^2  \sqrt{\lambda_{K^*}}}{3 (4 \pi)^5\, m_B^3}}\,,
\end{align}
and the $B \to K^*$ form factors $V$, $A_{1,2}$,  are defined as in
\cite{Bobeth:2010wg,Ball:2004rg}.

Finally, we also match the D-wave projection onto the $K_2^*$ contribution in zero-width approximation:
\begin{align}
\frac{d^2 \Gamma (D)}{d q^2  d \cos \theta_K }&=
   \int\! d p^2 \,\delta\! \left(p^2-m_{K_2^*}^2\right) \left.\frac{d^3 \Gamma}{d q^2 d p^2 d \cos \theta_K } \right|_D\nn\\ &=
   \frac{1}{12}|F_{0D}|^2\rho_1^-(1+3\cos^2 \theta_K-9\sin^2\theta_K\cos^2\theta_K)+
   3\left(|F_{\|D}|^2\rho_1^-+|F_{\perp D}|^2\rho_1^+\right) \cos^2\theta_K\sin^2\theta_K,\\
\frac{d  \Gamma (D)}{d q^2 } &  = \frac{2}{15}\left[\left(|F_{0D}|^2+6|F_{\|D}|^2\right)\rho_1^-+6|F_{\perp D}|^2\rho_1^+\right].
\end{align}
Matching onto Eq.~(10) of \cite{Li:2010ra} (LLW'10) yields (note that $\theta_K^{DHJS}=\pi-\theta_K^{LLW'10}$)
\begin{equation}
F_{0D} = \sqrt{15}A_0\, e^{i \delta_D}\,,\qquad F_{\parallel D} = \sqrt{\frac{5}{2}}A_{\parallel}\, e^{i \delta_D}\,,\qquad F_{\perp D} = -\sqrt{\frac{5}{2}}A_\perp\, e^{i \delta_D}\,,
\end{equation}
where we defined
\begin{align}
A_0 &= \mathcal N_{K_2^*} \frac{\sqrt{\lambda_{K_2^*}}}{\sqrt{24}\,m_B\,m_{K_2^*}^2\sqrt{q^2}}\left[\left(m_B^2-m_{K_2^*}^2-q^2\right)(m_B+m_{K_2^*})\tilde A_1-\frac{\lambda_{K_2^*}}{m_B+m_{K_2^*}}\tilde A_2\right]\,,\\
A_\parallel &= \mathcal N_{K_2^*} \frac{\sqrt{\lambda_{K_2^*}}}{2m_B\,m_{K_2^*}}(m_B+m_{K_2^*})\tilde A_1\,,\quad{\rm and}\\
A_\perp &= -\mathcal N_{K_2^*}\frac{\lambda_{K_2^*}}{2m_B\,m_{K_2^*}}\frac{\tilde V}{m_B+m_{K_2^*}}\,,
\end{align}
with the normalization factor
\begin{align}
  \label{eq:NK2star}
  {\mathcal N}_{K_2^*} & = G_F   V_{tb}^{}V_{ts}^{*} \alpha_e\,
    \sqrt{\frac{q^2  \sqrt{\lambda_{K_2^*}}}{3 (4 \pi)^5\, m_B^3}} \sqrt{\mathcal{B}(K_2^*\to K\pi)}\,.
\end{align}
For the definitions of the form factors in these equations, see~\cite{Li:2010ra}; note that we added a tilde to distinguish them from the ones in the $B\to K^*$ transition.

\end{document}